\documentclass[a4paper,aps,pra,showpacs,floatfix,superscriptaddress,twocolumn]{revtex4-1} %nofootinbib
\usepackage[osf,sc]{mathpazo}
\usepackage[colorlinks,breaklinks,linkcolor=magenta,citecolor=blue,urlcolor=blue]{hyperref}
\usepackage{amsmath,amssymb,amsthm,bm,mathtools,amsfonts,mathrsfs,bbm,dsfont}
\usepackage{physics}
\usepackage{graphicx}
\usepackage{color}
\usepackage{xcolor}
\usepackage{enumitem}

\newcommand{\id}{\ensuremath{\mathds{1}}}

\renewcommand{\ll}{\ensuremath{\bm{l}}}
\newcommand{\mm}{\ensuremath{\bm{m}}}

\newcommand{\HH}{\ensuremath{\mathcal{H}}}
\newcommand{\KK}{\ensuremath{\mathcal{K}}}
\renewcommand{\vec}[1]{\boldsymbol{#1}}

\newcommand{\bbC}{\mathbb{C}}
\newcommand{\bbR}{\mathbb{R}}
\newcommand{\sgn}{\mathrm{sign}}
\newcommand{\PS}[1]{P^{(#1)}}

\newtheorem{thm}{Theorem}

\newtheorem{lem}[thm]{Lemma}

\newtheorem{definition}[thm]{Definition}

\begin{document}
	\nonfrenchspacing
	
	\title{Indistinguishability of identical bosons from a quantum information theory perspective}
	
	\author{Matthias Englbrecht}
	\affiliation{Institute for Theoretical Physics, University of Innsbruck, Technikerstraße 21A, 6020 Innsbruck, Austria}
	\affiliation{Department of Physics, QAA, Technical University of Munich, James-Franck-Str. 1, D-85748 Garching, Germany}
	
	\author{Tristan Kraft}
	\affiliation{Institute for Theoretical Physics, University of Innsbruck, Technikerstraße 21A, 6020 Innsbruck, Austria}
	\affiliation{Department of Physics, QAA, Technical University of Munich, James-Franck-Str. 1, D-85748 Garching, Germany}

 \author{Christoph Dittel}

    \affiliation{Physikalisches Institut, Albert-Ludwigs-Universit\"at Freiburg, Hermann-Herder-Straße 3, 79104 Freiburg, Germany}
    \affiliation{EUCOR Centre for Quantum Science and Quantum Computing,
Albert-Ludwigs-Universit\"at Freiburg, Hermann-Herder-Straße 3, D-79104 Freiburg, Germany}
    \affiliation{Freiburg Institute for Advanced Studies, Albert-Ludwigs-Universit\"at Freiburg, Albertstraße 19, D-79104 Freiburg, Germany}

 \author{Andreas Buchleitner}
	\affiliation{Physikalisches Institut, Albert-Ludwigs-Universit\"at Freiburg, Hermann-Herder-Straße 3, 79104 Freiburg, Germany}
    \affiliation{EUCOR Centre for Quantum Science and Quantum Computing,
Albert-Ludwigs-Universit\"at Freiburg, Hermann-Herder-Straße 3, D-79104 Freiburg, Germany}
	
	\author{Geza Giedke}
	\affiliation{Donostia International Physics Center (DIPC), Paseo Manuel de Lardizabal 4, E-20018 San Sebasti\'an, Spain}
	\affiliation{IKERBASQUE, Basque Foundation for Science, Plaza Euskadi 5, E-48009 Bilbao, Spain}
	
	\author{Barbara Kraus}
	\affiliation{Institute for Theoretical Physics, University of Innsbruck, Technikerstraße 21A, 6020 Innsbruck, Austria}
	\affiliation{Department of Physics, QAA, Technical University of Munich, James-Franck-Str. 1, D-85748 Garching, Germany}
	
	\date{\today}
	
	\begin{abstract}
		Using tools from quantum information theory, we present a general theory of indistinguishability of identical bosons in experiments consisting of passive linear optics followed by particle number detection. Our results do neither rely on additional assumptions on the input state of the interferometer, such as, for instance, a fixed mode occupation, nor on any assumption on the degrees of freedom that potentially make the particles distinguishable. We identify the expectation value of the projector onto the $N$-particle symmetric subspace as an operationally meaningful measure of indistinguishability, and derive tight lower bounds on it that can be efficiently measured in experiments. Moreover, we present a consistent definition of perfect distinguishability and characterize the corresponding set of states. In particular, we show that these states are diagonal in the computational basis up to a permutationally invariant unitary. Moreover, we find that convex combinations of states that describe partially distinguishable and perfectly indistinguishable particles can lead to perfect distinguishability, which itself is not preserved under convex combinations.
	\end{abstract}
	
	\maketitle

	{\it Introduction.---}Indistinguishability of identical particles plays an important role in interference experiments. Those include the famous Hong-Ou-Mandel (HOM) experiment~\cite{Hong1987}, quantum computing with passive linear optics and particle number detection~\cite{Knill2001}, or boson sampling~\cite{Aaronson2013}, just to mention a few. In case of two particles interfering on a beam-splitter, the HOM effect completely characterizes the indistinguishability of the particles. Beyond that, most works make additional assumptions on the state of the particles. In particular, due to their experimental relevance, fixed mode occupation on the input of an interferometer~\cite{Shchesnovich2015,Brunner2022}, and other subsets of states have been considered~\cite{Brod2019}. A general theory of indistinguishability, including operationally motivated measures thereof, is still missing. This is precisely the aim of this paper.
	
	Throughout this paper we consider bosons, i.e.,  indistinguishable particles whose total state is symmetric. In case experiments operate only on a subset of their degrees of freedom (dofs), the so-called external dofs, the identical particles can become (partially) distinguishable. Here, we consider experiments involving passive linear optics and particle number detection in the external dofs. All other dofs, the so-called internal dofs, remain unchanged and unobserved in any such experiment. Our approach, which uses tools of quantum information theory, will thus only consider properties of arbitrary external density matrices, for which we show that the experiments described above are tomographically complete. That is, any two different reduced external density matrices can be distinguished by the experiments of the form described above.
    
    First, we present a definition of indistinguishability for arbitrary states with fixed total particle number, that is completely agnostic with respect to the internal dofs. More precisely, we will call particles indistinguishable, if, externally, they still behave like bosons, i.e., if their external density matrix has support only on the symmetric subspace. We identify the expectation value of the projector onto the $N-$particle (totally) symmetric subspace $\PS{N}$ as a measure of indistinguishability, and discuss its operational meaning. Moreover, we derive tight upper and lower bounds on the expectation value of $\PS{N}$ which can be efficiently measured in experiments. Then, we give a consistent definition of perfect distinguishability, that maximally violates the definition of perfectly indistinguishable particles. We provide a simple characterization of all states that describe perfectly distinguishable particles using tools from representation theory. Furthermore, we show that perfect distinguishability is not preserved under convex combinations and study the effect of convex combinations in the context of distinguishability. Finally, we briefly discuss the role of indistinguishability and entanglement in the emergence of many-body interference~\cite{dufour2020many,brunner2023many}.
	
\begin{figure}[b!]
		\centering
		\includegraphics[width=\columnwidth]{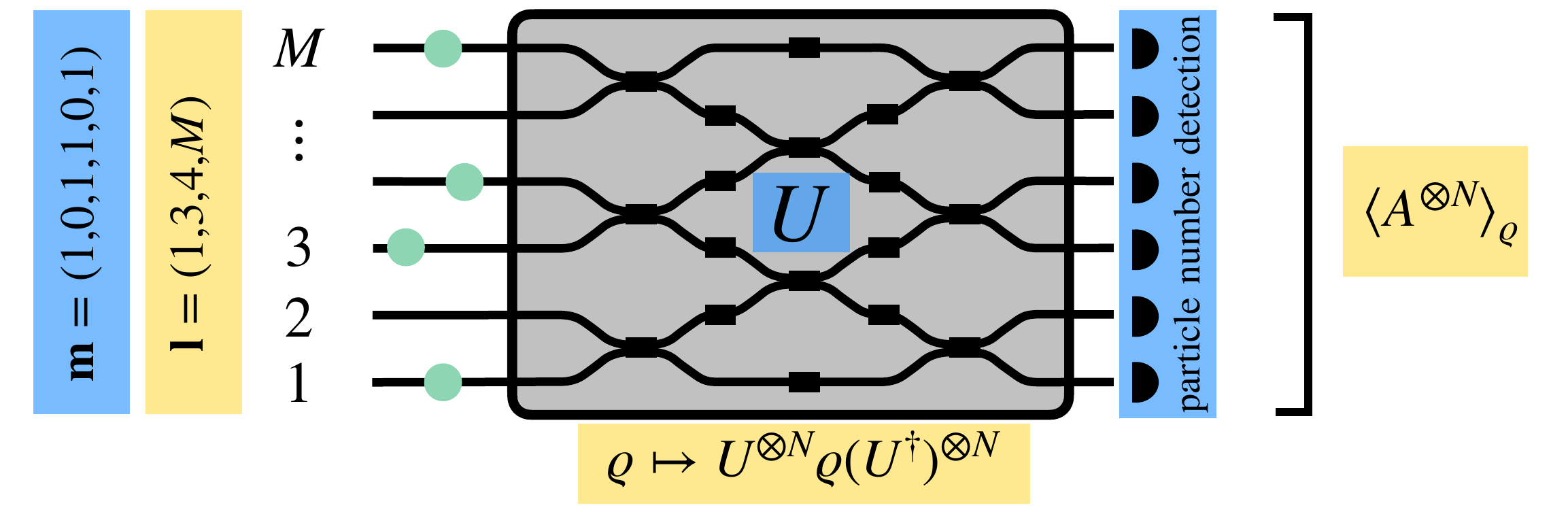}
		\caption{The experimental setup we consider: $N$ photons are injected into a linear optical interferometer followed by particle number detection on the output modes. Description in first quantization (orange), and description in second quantization (blue), as explained in the main text.}
		\label{fig:interferometer}
	\end{figure}

	{\it Notation \& Preliminaries.---} We consider a fixed number $N$ of particles (bosons) in $M$ external modes of a passive linear interferometer (see Fig.~\ref{fig:interferometer}). As it is usually done, we divide the single-particle Hilbert space into its external and internal dofs, i.e., $\HH=\HH_{\rm ext}\otimes \HH_{\rm int}$. The Hilbert space $\HH_{\rm ext}$ is associated to the external degrees of freedom which are the input modes of the linear interferometer. The Hilbert space $\HH_{\rm int}$ is associated to the internal dofs, which can neither be manipulated nor resolved in this experiment. For instance, these could include polarization, orbital angular momentum or time-of-arrival.

	In second quantization $a_{jk}^\dagger$ denotes the operator that creates a particle in the external mode $j$ and the internal mode $k$, and similarly for annihilation operators. We call the vectors $\mm=(m_1,\dots,m_M)$ the \emph{mode occupations}, meaning that mode $j\in\qty{1,\dots,M}$ is occupied by $m_j$ particles, and $\sum_j m_j =N$. The action of a linear interferometer transforms the creation operators to $b^\dagger_{ik}=\sum_j U_{ij}a_{jk}^\dagger$, where $U$ is unitary, and similarly for annihilation operators, thus acting only on the external dofs. The output modes of the interferometer are coupled to photon number detectors which, however, do not resolve any internal dofs, i.e., the measured observables are $M_j =\sum_k a_{jk}^\dagger a_{jk}$ in second quantization.
	
	In first quantization one defines \emph{mode assignment vectors} $\ll=(l_1,\dots,l_N) \in \mathcal{H}_{\rm ext}^N=(\mathbb{C}^M)^{\otimes N}$, with $l_k\in\qty{1,\dots,M}$ (computational basis) denoting the mode in which each particle $k$ is located (cf. Ref.~\cite{tichy2012many}). We have to account for the symmetrization of bosonic states. To that end, let us first define the action of the permutation group $S_N$ on the total Hilbert space. A permutation $\pi\in S_N$ acts on the mode assignment vectors as $\pi\ket{\ll}=\ket*{l_{\pi(1)},\dots,l_{\pi(N)}}$, where, with a slight abuse of notation, $\pi$ denotes elements in $S_N$ and their unitary representation on $\HH^{N}_{\rm ext}$, cf. App.~\ref{app:group_intro_and_reps}. The action on the internal Hilbert space is defined similarly. The overall state must be symmetric under permutation of particles, i.e., $\pi\otimes \tilde{\pi}\ket{\psi}=\ket{\psi}\, ,\forall \pi\in S_N$, where $\pi$ ($\tilde{\pi}$) denotes the representation of a permutation $\pi$ in the external (internal) Hilbert space, respectively. The projector onto the $N-$particle symmetric subspace of $\HH^N$ is given by
	\begin{equation}\label{eq:PS}
		P^{(N)}_\text{tot} = \frac{1}{N!}\sum_{\pi\in S_N} \pi\otimes\Tilde{\pi},
	\end{equation}
	and $\PS{N} = \frac{1}{N!}\sum_{\pi} \pi$ for $\HH^N_{\rm ext}$, and similarly for $\HH^N_{\rm int}$. Note, that these are all real and orthogonal projectors for which $P^2=P=P^T$.
	
	As a simple example we consider the Fock-state with a single, i.e., fixed, external mode occupation $(2,0)$, and internal occupation $(1,1)$. In first quantization this state is $a_{1,1}^\dagger a_{1,2}^\dagger\ket{0}\propto\ket{11}_{\rm ext}\otimes (\ket{12}+\ket{21})_{\rm int}$. In case the system is entangled with auxiliary (non-bosonic) dofs, the symmetry of the state reads $\pi\otimes \tilde{\pi}\otimes \id_{\rm aux}\ket{\psi}=\ket{\psi}\, ,\forall \pi\in S_N$, and thus for bosonic mixed states one has $P_\text{tot}^{(N)}\varrho=\varrho$.
	
	From the permutation symmetry of bosonic states it directly follows that their external density matrices $\varrho_{\rm ext}=\tr_{\rm int}[\ketbra{\psi}]$ have to be permutation invariant, i.e., $\pi\varrho_{\rm ext}\pi^\dagger=\varrho_{\rm ext}\,,\forall \pi\in S_N$. We denote the set of all permutation invariant states on $\mathcal{H}_{\rm ext}$ as $\mathrm{Inv}(\mathcal{H}_\text{ext})=\{ \varrho_\text{ext}\ :\pi \varrho_\text{ext} \pi^\dagger = \varrho_\text{ext}\ ,\forall \pi\in S_N,\ \varrho_\text{ext}\geq 0,\ \tr[\varrho_\text{ext}]=1\}$. Due to Schur-Weyl duality any $\varrho\in {\rm Inv}(\mathcal{H})$ can be written as a direct sum $\varrho=\bigoplus_{\bm{\lambda}} p_{\bm{\lambda}} \varrho_{\bm{\lambda}}\otimes \id/d_{\bm{\lambda}}$, where $\bm{\lambda}=(\lambda_1,\lambda_2,\dots, \lambda_M)$ is a partition of $N$ into $M$ decreasingly ordered and non-negative integers,  $\{p_{\bm{\lambda}}\}$ is a probability distribution, and $\varrho_{\bm{\lambda}}$ are density matrices supported on invariant subspaces [cf. Eq.~\eqref{eq:SWdec}].
	
	One can show that any permutation invariant state admits a permutation symmetric purification~\cite{Renner2005}, see App.~\ref{app:purifications}. Crucially, the purification is not unique. For instance, one can see that the state $\ket{11}_{\rm ext}\otimes \ket{11}_{\rm int}$ and the state discussed above lead to the same $\varrho_{\rm ext}$.
	
	Since the interferometer only acts on external dofs it corresponds, in first quantization, to the transformation 
	\begin{equation}
		\varrho_{\rm ext} \mapsto U^{\otimes N}\varrho_{\rm ext}(U^{\dagger})^{\otimes N}
	\end{equation}
	with a single-particle unitary $U$. Furthermore, the subsequent measurement (see Fig.~\ref{fig:interferometer}) projects the state onto a single mode occupation. In first quantization, this measurement is effectively described by the projector onto a computational state, i.e., the corresponding mode assignment vector, and all its permutations (see App.~\ref{app:red_to_qudits}).
	
	We show in App.~\ref{app:tomography}, using similar arguments as in the case of qubits in Ref.~\cite{Toth2010}, that for the external density matrix $\varrho_{\rm ext}\in {\rm Inv}(\mathcal{H}_\text{ext})$ such experiments are tomographically complete. Therefore, if particles have the same external density matrix they show the same behaviour in any of these experiments, and if they do not there exists an experiment where the two states lead to different outcome statistics.
	
	{\it Indistinguishable Particles.---} As mentioned in the introduction we call particles indistinguishable if, externally, particles behave like bosons. More precisely, identical particles described by a state $\varrho_{\rm ext}\in {\rm Inv}(\mathcal{H}_\text{ext})$ are called perfectly indistinguishable if
    \begin{equation}
	     \PS{N}\varrho_{\rm ext}=\varrho_{\rm ext}.
    \end{equation}
	Some remarks are in order. (i) Given a $\varrho_{\rm ext}$, our definition of perfectly indistinguishable particles is independent of the purification and only refers to the accessible dofs of the particles. (ii) In case the photons are entangled with auxiliary, i.e., non-bosonic dofs, we define them to be perfectly indistinguishable if $\pi \otimes\id_{\rm int}\otimes\id_{\rm aux}\ket{\psi}=\ket{\psi}\,, \forall \pi\in S_N$. Hence, this is equivalent to the condition above. As from now on we will be only concerned with properties of the external density matrix, we will use the notation $\varrho$ for $\varrho_{\rm ext}$.
	
	This definition of indistinguishability suggests the following quantifier for indistinguishability
	\begin{equation}\label{eq:PSN}
		p_N \coloneqq \expval*{\PS{N}}_{\varrho_{\rm ext}}= \mathrm{tr}[\PS{N}\varrho_{\rm ext}].
	\end{equation}
	Clearly, $p_N=1$ for perfectly indistinguishable particles, and $p_N=0$ for states that have no support on the symmetric subspace. Recall that any bosonic state can be written as a direct sum  of one part, which is in the symmetric subspace, whose weight is measured by $p_N$, and its orthogonal complement. Note that this measure coincides with measures discussed in Refs.~\cite{Tichy2015,Shchesnovich2015trace,Brunner2022,Dittel2021}, where certain subsets of states have been considered (for more details, see App.~\ref{app:relation}). In these special cases other measures for indistinguishability have been considered, see, e.g., Refs.~\cite{Shchesnovich2015,Dittel2019,Karczewski2019,Walschaers2020,Dittel2021,Brunner2022}. 
	
	Another measure can be obtained from the trace distance between two states, which is essentially the maximum probability with which one can discriminate the two states. It is defined by $D_{\rm tr}(\sigma, \varrho)=\frac{1}{2}\tr[\abs{\sigma - \varrho}]$, with $\abs{X}=\sqrt{X^\dagger X}$. One can show that $D_{\rm tr}(\varrho_{\rm ind},\varrho) =1-p_N$, where $\varrho_{\rm ind}\equiv \PS{N}\varrho\PS{N}/p_N$ denotes the indistinguishable part of $\varrho$~\footnote{Using that $D_{\rm tr}(\sigma, \varrho)=\mathrm{tr}[(\sigma - \varrho)_+]$, where $(X)_+$ denotes the positive part of $X$ and the fact that, due to Schur-Weyl duality (see App.~\ref{app_schurweyl}), for any permutation-invariant state, such as the indistinguishable part $\varrho_{\rm ind}\equiv \PS{N}\varrho\PS{N}/p_N$ of $\rho$, there is no coherence between the symmetric subspace and its orthogonal complement, we have $D_{\rm tr}(\varrho_{\rm ind},\varrho) = \mathrm{tr}[\varrho_{\rm ind}-\PS{N} \varrho\PS{N}]=1-p_N$.}. Furthermore, it is a well known fact that the trace distance between two states provides an upper bound to the total variational distance between the probability distributions that are obtained by performing measurements on the two states. This provides the following lower bound on $1-p_N$.
	\begin{thm}\label{thm:tracedist}
		It holds that $D(r_k, q_k)=\frac{1}{2}\sum_k \abs{r_k-q_k}\leq 1-p_N$, where $r_k=\mathrm{tr}[M_k\varrho_{\rm ind}]$, and $q_k=\tr[M_k\varrho]$. In particular, equality holds if the left hand side is maximized over all measurements $\{M_k\}$.
	\end{thm}
	This generalizes the result of Ref.~\cite{Shchesnovich2015trace} from fixed mode occupations to arbitrary states. To the best of our knowledge, there does not exist an efficient way to estimate $p_N$. In Ref.~\cite{Shchesnovich2015trace} it was shown to be related to the probability of finding all bosons bunched in a single mode, but for instance for states with at most one particle per mode this probability decreases exponentially in the number of particles $N$ \footnote{Note also that pairwise HOM experiments between all particles for states $\varrho$ with exactly 0 or 1 particles in each of the modes cannot be used to determine $P_N$.}. Also other approaches to quantify indistinguishability~\cite{Pont2022} rely on measuring probabilities that become exponentially small with $N$. Here we show how to obtain optimal lower bounds on $p_N$ only as a function of $p_2=\mathrm{tr}[(P^{(2)}\otimes\mathds{1}) \varrho]$ [cf. Eq.~\eqref{eq:PSN}], which, as we show, can be efficiently measured~\footnote{Note that in Refs.~\cite{brunmsc,Brunner2022}, $p_2$ has been used to derive an upper bound on $p_N$ and it has been related to another notion of indistinguishability~\cite{Brod2019}. Note that these results only apply to a particular subset of so called externally separable states. For externally separable states (with fixed total particle number) the expectation values $p_k$ agree with the mean coherence, which is the quantity studied in Refs.~\cite{brunner2020many,Brunner2022}. For more details and the definitions of external separability and the mean coherence we refer the reader to Refs.~\cite{brunner2020many,Brunner2022} for more details)}. 
	
	\begin{figure}[t!]
		\centering
		\includegraphics[width=0.9\columnwidth]{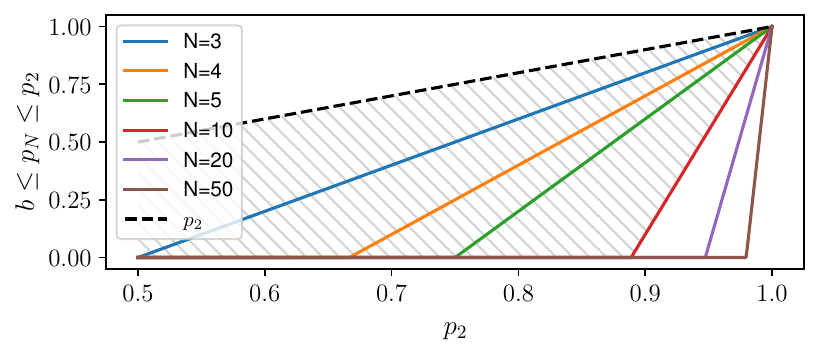}
		\caption{Tight upper and lower bounds on $p_N$, for different $N$, as a function of $p_2$, cf. Eq.~\eqref{eq:bounds}. The dashed area are all states for $N=10$ that are compatible with the respective value for $p_2$. A large value of $p_2$ implies high indistinguishability. For a given particle number $N$, $p_2$ provides a nontrivial lower bound on $p_N$ if $p_2\ge (N-2)/(N-1)$. That is for a larger number of particles, higher values of $p_2$ are required to obtain a nontrivial lower bound on $p_N$.}
		\label{fig:bounds}
	\end{figure}
	
	{\it Optimal bounds on $p_N$ from $p_2$.---} For $\varrho\in {\rm Inv}(\mathcal{H}_\text{ext})$ a $N-$particle state, it is straightforward to show that $p_2=1$ implies $p_N =1$. As the following theorem shows if $p_2$ is sufficiently large, one can also derive tight lower bounds on $p_N$. The reason for that is that a large $P_2$ imposes strong conditions on any state in $ {\rm Inv}(\mathcal{H}_\text{ext})$.
	\begin{thm}\label{thm:bounds}
		It holds that $b\leq  p_N\leq p_2$ (cf. Fig.~\ref{fig:bounds}), where
		\begin{eqnarray}\label{eq:bounds}
			b = \begin{cases}
				0 &\text{if } \, 0\leq p_2\leq \frac{N-2}{N-1},\\
				p_2(N-1)-(N-2) &\text{if }\, \frac{N-2}{N-1}\leq p_2\leq 1,
			\end{cases}
		\end{eqnarray}
		Hence, if $p_2$ is sufficiently large one obtains strong lower bounds on $p_N$ in terms of $p_2$. Moreover, these bounds are tight.
	\end{thm}
	Let us first outline how the lower bound can be derived (cf. App.~\ref{app:bounds} for a detailed proof). Finding a tight lower bound on $p_N$ can be cast as the following optimization problem
	\begin{align}\label{eq:sdp}
		b&= \min\left\{ \mathrm{tr}[\PS{N}\varrho] :\varrho\in \mathrm{Inv}(\mathcal{H}),\ \mathrm{tr}[\PS{2}\varrho]=p_2\right\}.
	\end{align}
	As $\varrho\in \text{Inv}(\mathcal{H})$, one can rewrite the last condition as $\tr[T\varrho]=c$, where $T=2/[N(N-1)]\sum_{i\neq j}\pi_{ij}$ is a sum over all transpositions in $S_N$, and thus hermitian, and $c=2p_2-1\in \qty[0,1]$. As $[\varrho,T]=0$, the operators $\varrho$ and $T$, can be simultaneously diagonalized. Using the decomposition of $\varrho\in\text{Inv}(\HH_{\rm ext})$ as $\varrho=\bigoplus_{\bm{\lambda}} p_{\bm{\lambda}} \varrho_{\bm{\lambda}}\otimes \id/d_{\bm{\lambda}}$ leads to an easily solvable linear optimization problem with Lagrange multipliers. One can obtain an upper bound by simply replacing the minimization in Eq.~\eqref{eq:sdp} by a maximization. Tightness can be shown by constructing states that saturate the bounds.
	
	To see that $p_2$ can be efficiently measured, we observe that $\expval*{(A^{\otimes 2}\otimes\id^{\otimes (N-2)})_{PI}}= {N\choose 2}\expval{A\otimes A}_{\varrho_2}$ (cf.~Ref.~\cite{Toth2010}), where $(\cdot)_{PI}=\sum_\pi \pi\cdot\pi^\dagger$, $\varrho_2$ is the two-particle reduced state, and the last equality holds due to the permutation invariance of $\varrho$. As $\PS{2}$ is permutation invariant, it can be expressed as a linear combination of ${M^2+1 \choose M^2-1}\propto M^4$ many operators of the form $A\otimes A$ (see App.~\ref{app:tomography}). Thus, $p_2$ can be estimated efficiently~\footnote{For particular subsets of input states, $p_2$ can even be determined with less measurements. It is straightforward to show that for input states with exactly 0 or 1 particles in each external mode, the measurement outcomes from all $N(N-1)/2$ pairwise HOM experiments between all particles can be used to compute $p_2$.}. Therefore, if $N$ particles are sufficiently indistinguishable, this can be verified efficiently. Note that similar techniques as above can be used to derive bounds on $p_N$ that result from knowledge of $p_k$ for any $k\ge 3$. One obtains a hierarchy of bounds providing stronger bounds on $p_N$.
 
 Note that for input states with $p_N=1$ and $m=(1,1\ldots,1)$ (Fock states) it has been shown that boson sampling cannot be classically simulated efficiently~\cite{Aaronson2013}, provided that some assumptions from complexity theory are true. For states with $k$ orthogonal internal dofs, e.g., for the state $\sum_{\pi\in S_N}\pi\ket{12345}_{\rm ext}\otimes\pi\ket{11123}_{\rm int}$ with $k=3$, boson sampling is as hard as for $N-k+1$ indistinguishable particles. For these states, $\rho_{\rm ext}$ has specific weights in the various invariant subspaces~\cite{dufour2020many}. This suggests that these weights (or functions thereof) might allow us to gain further information about the indistinguishability properties of the particles described by these states~\footnote{An instance of a measure which does not only consider $p_N$ is the purity of $\varrho_{ex}$ as studied in Ref.~\cite{brunner2023many}.}.

	{\it Perfectly distinguishable particles.---} Inspired by the definition of perfect indistinguishability one could define that a state describes perfectly distinguishable particles if $\tr[\pi\varrho_{\rm ext}]=0\, \forall\pi\neq\id$. However, there exist states with a fixed mode occupation that fulfill these conditions but fail to fulfill other definitions of perfect distinguishability in the literature, see App.~\ref{app:motivation}. Slightly adjusting the definition avoids this contradiction. More precisely, we call identical particles described by a state with spectral decomposition $\varrho= \sum_{k}q_k P_k$ and $q_k>0$ perfectly distinguishable if for any projector $P_k$ on its eigenspaces $V_k$ it holds that
	\begin{eqnarray}\label{eq:distsubspace}
		\tr[\pi P_k] = \begin{cases}
			\dim(V_k) &\text{if } \pi=\text{id},\\
			0 \quad &\text{otherwise}.
		\end{cases}
	\end{eqnarray}
	Let us note that our definition of perfect distinguishability still partly differs from definitions from literature for fixed mode occupation (see App.~\ref{app:comparison_literature} for details).

		Hence, the eigenspaces of a state describing perfectly distinguishable particles are very restricted. In fact, as we will show next, they are spanned, up to a permutationally invariant unitary by computational basis vectors with distinct entries  and their permutations. To show that, we first characterize all projectors $P_k$ that fulfill Eq.~\eqref{eq:distsubspace}. If we consider the unitary representation $\pi$ of the permutation group on $(\mathbb{C}^M)^{\otimes N}$, and restrict it to the subspace $V_k\subset (\mathbb{C}^M)^{\otimes N}$ we obtain a unitary representation $\pi_k=\pi|_{V_k}=P_k\pi P_k$ of the permutation group, since $V_k$ is an invariant subspace (see App.~\ref{app:representations}). It is well known that representations of finite groups are completely classified by their \emph{character} $\chi[\pi_k] \equiv \tr[\pi_k]$, and Eq.~\eqref{eq:distsubspace} defines a character. This allows us to prove the following result in App.~\ref{app:distinguishability}.
		\begin{thm}\label{thm:distinguishable}
			A state $\varrho = \sum_{k}q_k P_k$ is perfectly distinguishable iff there exists a permutation invariant unitary $U$ such that each $\tilde{P}_k=UP_kU^\dagger$ is of the form $\tilde{P}_k=\oplus_{i_k} \tilde{P}_{k,i_k}$, where $\tilde{P}_{k,i_k}=\sum_\pi \pi \ketbra{l_1,\dots,l_N} \pi^\dagger$, with  $l_1<\dots <l_N$ and $(l_1,\dots,l_N)$ such that the projectors $\tilde{P}_{k,i_k}$ are all mutually orthogonal for $k$ and $i_k$.
		\end{thm}
		The maximum number of orthogonal subspaces, onto which $\tilde{P}_{k,i_k}$ project, is $M \choose N$ which is the dimension of the anti-symmetric subspace (see App.~\ref{app:distinguishability}). Therefore, states that describe perfectly distinguishable particles can only exist if $M\geq N$.
		
\begin{figure}[t!]
			\centering
			\includegraphics[width=\columnwidth]{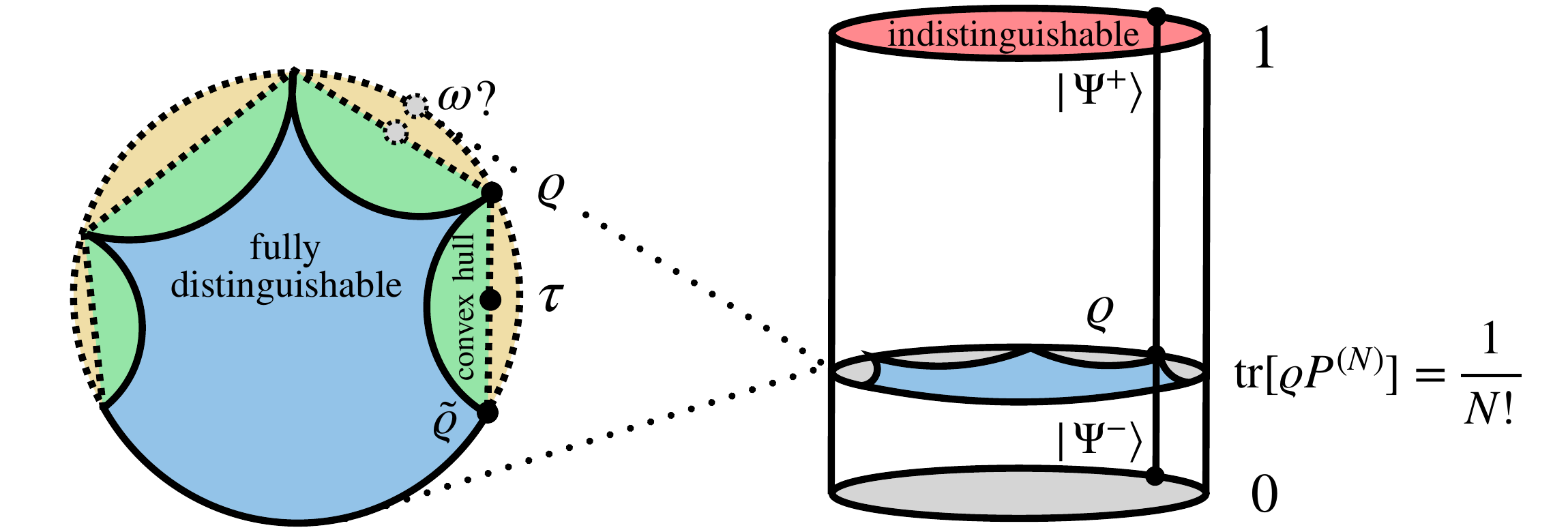}
			\caption{(right) The set of $N-$ particle external density matrices according to their weight on the symmetric subspace. The subset fulfilling Eq.~\eqref{eq:convexhull} (left), contains the non-convex set of perfectly distinguishable states (blue). Mixing perfectly distinguishable states $\varrho$ and $\tilde{\varrho}$ results in a state $\tau$ not satisfying Theorem~\ref{thm:distinguishable}. It is unclear if there exist states $\omega$ (yellow region), outside the convex hull of perfectly distinguishable states, that satisfy Eq.~\eqref{eq:convexhull}. Moreover, it is possible to mix perfectly and partially indistinguishable states to obtain perfectly distinguishable states. For instance, $\ketbra{\Psi^+}+\ketbra{\Psi^-}=\ketbra{12}+\ketbra{21}$ is perfectly distinguishable.}\label{fig:convex}
		\end{figure}

		For instance, the fixed mode occupation state $\varrho=\sum_{\pi} \pi\ketbra{1\dots N}\pi^\dagger$ is perfectly distinguishable, and the same is true for any state $\tilde{\varrho} = U^{\otimes N}\varrho (U^{\dagger})^{\otimes N}$. However, in case the states $\rho$ and $\tilde{\varrho}$ are not orthogonal, their convex mixture $\tau = \frac{1}{2}(\varrho + \tilde{\varrho})$ does not fulfill Theorem~\ref{thm:distinguishable}, for a suitably chosen $U$. This shows that the set of perfectly distinguishable states is not convex, cf. Fig.~\ref{fig:convex}. Moreover, our definition of perfect distinguishability implies that for perfectly distinguishable states 
		\begin{equation}\label{eq:convexhull}
			\mathrm{tr}[\PS{N}\varrho]=\frac{1}{N!},
		\end{equation}
		but the converse is not true due to the existence of the state $\tau$, which satisfies this condition. However, it is currently unclear if the convex set defined by Eq.~\eqref{eq:convexhull}, is larger than the convex hull of perfectly distinguishable states, cf. Fig.~\ref{fig:convex}. This shows, that in general, the expectation value of $\PS{N}$ is not a good quantifier for distinguishability, except for a fixed mode occupation~(see, e.g.,~\cite{Shchesnovich2015,Brunner2022}).
		
        Let us finally mention that it is clear that not all indistinguishable particles show many-body interference. A simple example is the state $\varrho_{\rm ext}=\ketbra{1\dots 1}$, as this shows the same interference pattern as $N$ independent particles (MBI, cf. App.~\ref{app:relation}).
		Moreover, indistinguishability is invariant under permutation-invariant unitaries, whereas interferometers implement $U^{\otimes N}$, which is a much smaller set. This suggests, that other properties of $\varrho_{\rm ext}$ are better suited to quantify the amount of MBI, such as the multiparticle entanglement of $\varrho_{\rm ext}$. Indeed, one notices that entanglement among the $N$ $M$--level systems is invariant under $U^{\otimes N}$, and states like $\varrho_{\rm ext}=\ketbra{1\dots 1}$ or perfectly distinguishable states contain no entanglement, whereas Fock states are multipartite entangled (in the sense of being non-separable due to their symmetrization).
		
		{\it Conclusion.---} We have introduced a general theory of indistinguishability of identical particles that does not rely on additional assumptions on the internal dofs, that are inaccessible in experiments. We have shown  that the external density matrix  contains precisely the information accessible about the particles in interference experiments, i.e., linear interferometers followed by particle number detection. We have put forward definitions of indistinguishability and distinguishability based on the reduced external state, whereby the definition of perfect distinguishability maximally violates the condition for perfect indistinguishability. We also characterized both sets of states using the representation theory of the symmetric group. Moreover, we have shown that for arbitrary states, the expectation value of the projector onto the symmetric subspace is an operationally meaningful measure of indistinguishability. Furthermore, we have derived optimal upper and lower bounds on our indistinguishability measure that can be obtained from the indistinguishability of the two-particle reduced state and hence efficiently measured.
		
		Our work also opens many interesting future research directions. For instance, it would be appealing to study how the weight of a state in invariant subspaces other than the symmetric one~\cite{brunner2023many} can be taken into account to define more fine-grained measures than $p_N$~\cite{Note5}. Moreover, one could imagine using tools from multipartite entanglement theory to study the role of entanglement of $\varrho_{\rm ext}$ in the emergence of MBI and, in particular, in a possible resource theory of indistinguishability.
  
		{\it Acknowledgements.---} We thank J. Ignacio Cirac, Ingo Roth, and Benoît Vermersch for interesting  discussions. ME, TK, and BK  acknowledge financial support from the Austrian Science Fund (FWF)  through the grants SFB BeyondC (Grant No. F7107-N38), P 32273-N27 (Stand-Alone Project), and DK-ALM W1259-N27. ME and BK acknowledge the BMW endowment fund.  GG acknowledges funding by the European Union NextGenerationEU/PRTR-C17.I1, as well as by the IKUR Strategy under the collaboration agreement between Ikerbasque Foundation and DIPC on behalf of the Department of Education of the Basque Government. C.D. acknowledges the Georg H. Endress Foundation for support and the Freiburg Institute for Advanced Studies for a FRIAS Junior Fellowship

		\onecolumngrid
		
		\clearpage
		
		\appendix
		
		\section{Tomographic completeness of interference experiments}\label{app:tomography}

		In this section, we show that interference experiments are tomographically complete as claimed in the main text. The setup of such experiments is shown in Fig.~\ref{fig:interferometer}. For readability, let us recall the precise statement again.
		
		\begin{thm}\label{thm:mm_settings}
			Let $\varrho\in\mathrm{Inv}((\bbC^M)^{\otimes N})$. Then, $\varrho$ can be uniquely characterized by the values $\braket{\vec{b}|(U^{\otimes N})^\dagger \varrho U^{\otimes N}}{\vec{b}}$ for all computational basis states $\ket{\vec{b}}$ and sufficiently many, randomly chosen unitaries $U\in U(M)$. That is $\varrho$ can be characterized by interference experiments. Stated differently, interference experiments are tomographically complete for $\varrho\in\mathrm{Inv}((\bbC^M)^{\otimes N})$.
		\end{thm}
		Note that for the case $M=2$, Theorem~\ref{thm:mm_settings} is already shown in Refs.~\cite{Toth2010,Kirk2022}.
		\begin{proof}
			Let $L$ be the subspace of $\mathcal{B}((\bbC^M)^{\otimes N})$ formed by operators $O\in \mathcal{B}((\bbC^M)^{\otimes N})$ such that $\pi O=O \pi$ for all $\pi\in S_N$. It is straightforward to verify that 
			\begin{equation}
				\dim(L)={
					N+M^2-1\choose M^2-1}.
			\end{equation}
			To show the theorem we show that there exists a basis of $L$ consisting of operators of the form $A^{\otimes N}$, where $A\in\mathcal{B}(\bbC^M)$ is hermitian, and that random choices for the operators $A^{\otimes N}$ yield such a basis. As the tools specified in the theorem allow one to determine the expectation values of such operators, $\varrho$ can be reconstructed. To determine the expectation value of an operator $A^{\otimes N}$, choose $U$ such that it diagonalizes the observable $A$. Then $\left< A^{\otimes N}\right>$ can be obtained from a computational basis measurement of $U^{\otimes N}\varrho (U^{\otimes N})^\dagger$ together with classical post-processing.
			
			Let us show that there exists a set of operators $\{A_i^{\otimes N}\}$ with $A_i\in\mathcal{B}(\bbC^M)$ hermitian such that it forms a basis of $L$. To this end, we choose a basis $G_1,\ldots G_{M^2}$ for the hermitian matrices in $\mathcal{B}(\bbC^M)$. Let us expand a hermitian $A\in \mathcal{B}(\bbC^M)$ in the above basis as $A=\sum_j \lambda_j G_j$, where $\lambda_j\in \bbR$ for all $j$. To express the $N$ fold tensor product $A^{\otimes N}$ with respect to the same basis, let us introduce some notation. For any $\vec{n}=(n_1,\ldots, n_{M^2}) \in\{1,\ldots, N\}^{M^2}$, we denote by $J(\vec{n})$ the operator
			
			\begin{equation}
				J(\vec{n})=\frac{1}{\prod_j n_j!}\sum_{\pi\in S_N} \pi\left(\bigotimes_{m=1}^{M^2}G_m^{\otimes n_m}\right)\pi^\dagger.
			\end{equation}
			Using this notation, the $N$ fold tensor product $A^{\otimes N}$ can be written as
			
			\begin{equation}
				A^{\otimes N}=\sum_{\substack{n_1,\ldots n_{M^2}=0\\\sum_j n_j=N}}^N\prod_{p=1}^{M^2}\lambda_p^{n_p}J(\vec{n}).
			\end{equation}
			
			As $G_1,\ldots G_{M^2}$ is a basis for $\mathcal{B}(\bbC^M)$, the operators $\{J(\vec{n})\}$ form a basis of the space $L$. We have to show that through suitable choices of hermitian operators $A_i\in \mathcal{B}(\bbC^M)$ we obtain $M^2$ independent operators $\{A_i^{\otimes N}\}$. This is equivalent to showing that the vectors of coefficients $\{\boldsymbol{\Lambda}(i)\in\bbR^{\dim(L)}\}_i$ of the operators $A_i^{\otimes N}$ with respect to the basis $\{J(\vec{n})\}$ can be chosen to be linearly independent. The vector of coefficients of $A_i^{\otimes N}$ can be regarded as a vector of monomials in the variables $\{\lambda_j^{(i)}\}_j$. For instance, for the operator $A_i=\sum_j \lambda_j^{(i)} G_j$, the component of the vector that corresponds to the basis element $G_1^{\otimes N}$ is $(\lambda_1^{(i)})^N$. In order to see that it is possible to choose operators $A_i$ such that $\{\boldsymbol{\Lambda}(i)\in\bbR^{\dim(L)}\}_i$ are independent, consider the following argument: Let us arrange the vectors $\{\boldsymbol{\Lambda}(i)\in\bbR^{\dim(L)}\}_i$ for $M^2$ different matrices $A_i^{\otimes N}$ in a matrix as $M=(\boldsymbol{\Lambda}[1],\ldots,\boldsymbol{\Lambda}[\dim(L)])$. The operators $\{A_i^{\otimes N}\}$ are linearly independent if $\det(M)\neq 0$. The determinant $\det(M)$ is a polynomial in the variables $\{\lambda_j^{(i)}\}_{i,j}$. As within a column all monomials are different, and the variables between different columns are not the same, it follows that $\det(M)$ is not equal to the $0$ polynomial. Therefore, $\det(M)\neq 0$ for all but a zero measure subset of choices for $\lambda_j^{(i)}\in\bbR^{k}$.
		\end{proof}
		
		That is applying $U^{\otimes N}$ to $\varrho$ for sufficiently many $U$ and measuring the resulting state in the computational basis is tomographically complete for permutation invariant states $\varrho$, i.e., for any $\varrho,\sigma\in \text{Inv}(\HH)$ with $\varrho\neq \sigma$ there exists a $U\in U(M)$ and a computational basis state $\vec{b}$ such that $\braket{\vec{b}|(U^{\otimes N})^{\dagger}\varrho U^{\otimes N}}{\vec{b}}\neq \braket{\vec{b}|(U^{\otimes N})^{\dagger}\sigma U^{\otimes N}}{\vec{b}}$. Note that it follows from the above proof that $K(M,N)$ different, randomly chosen unitaries $U\in U(M)$ suffice to determine $\varrho$, where
		\begin{equation}
			K(M,N)={
				N+M^2-1\choose
				M^2-1}.
		\end{equation}
		Note however that this number is not optimal, as for a single choice of $U$ knowledge of all probabilities of all computational basis states allows one to determine more the expectation value of any operator $A^{\otimes N}$ diagonalized by $U^{\otimes N}$.
		
		 Another consequence of the tomographical completeness is the following. For two particles, for which the internal state of each of the particles is well defined, the reduced density matrix $\varrho_{\rm ext}$ only depends on the overlap between the (possibly mixed) internal dofs. Thus, no interference experiment involving linear optics and particle-number detection can separate distinguishability due to entanglement with auxiliary dofs from distinguishability due to only partially overlapping internal pure states, as both can lead to the same reduced state. For more particles, the fact that the symmetry constraint affects only the internal and external dofs, leaves some room to differentiate between the effect of the entanglement between the external dofs and the internal or auxiliary, non-bosonic, dofs on the reduced state $\varrho_{\rm ext}$.

   Lastly, let us remark that it follows from Theorem~\ref{thm:mm_settings} that any device that is able to implement boson sampling~\cite{Aaronson2013} can also be used to obtained so called classical shadows of the input state. Classical shadows~\cite{huang2020predicting,elben2020mixed} are classical snapshots of the input state that can be used to determine the expectation value of many observables efficiently via classical post-pressing. That this is indeed the case for a boson sampler follows from the fact that any such device is able to implement Haar random unitaries and according to Theorem~\ref{thm:mm_settings} already finitely many random untaries are sufficient for the experiment to be tomographically complete. The latter property is precisely what is required to obtain classical shadows.

		\section{A brief introduction to finite groups and their representations}\label{app:group_intro_and_reps}
		In this appendix we give a brief introduction to finite groups and their representations. For more details we refer to Ref.~\cite{Simon1996}. A group is a set $G$ together with a map $\cdot: G\times G\rightarrow G$, the group operation, such that the following properties are satisfied: (i) There exists an identity element $ e\in G$ such that for all elements $g\in G$ it holds that $e\cdot g=g\cdot e=g$. (ii) For all elements $g\in G$ there exists an inverse element $g^{-1}\in G$ such that $g g^{-1}=g^{-1}g=e$. And finally, (iii) the group operation is associative, i.e., $a\cdot (b\cdot c)=(a\cdot b)\cdot c$ for all $a,b,c\in G$. The number of elements $\#(G)$ of a group $G$ is called the order of the group and we write it as $|G|$.
		
		An important substructure of groups are the so-called conjugacy classes. The conjugacy class of an element $g\in G$ is the set $\{hgh^{-1}|h\in G\}$. One can show that the conjugacy classes of any two elements $g_1,g_2\in G$ are either disjoint or equal. Conjugacy classes are particularly relevant for the study of the representations of a group, which we discuss below.
		
		A group homomorphism between two groups $G$ and $H$ is a map $E:G\rightarrow H$ that fulfills
		\begin{equation}
			E(gh)=E(g)E(h)
		\end{equation}
		for all $g,h\in G$. If a group homomorphism is bijective, it is called an isomorphism.
		
		\subsection{Representations of finite groups}\label{app:representations}
		A representation of a group $G$ is a group homomorphism $R:G\rightarrow GL(V)$, where $GL(V)$ are linear, invertible operators over some vector space $V$, the so called carrier space. One also says that $V$ carries a representation of $G$. The dimension of $V$ is called the dimension of the representation. Note that we are only concerned with finite-dimensional representations over complex vector spaces. Two representations $R_1$ and $R_2$ are said to be equivalent, written as $R_1\cong R_2$, if they are related by an isomorphism, i.e., there exists a similarity transformation $T$ such that $R_1(g)=TR_2(g)T^{-1}$. The representation $R$ is a unitary representation if $R(g)$ is unitary for all $g\in G$. Next, let us summarize some important theorems in the representation theory of finite groups.
		\begin{thm}
			Any representation is equivalent to a unitary representation.
		\end{thm}
		See, for instance, Theorem II.1.1. in Ref.~\cite{Simon1996}.
		\begin{thm}\label{thm_unitary}
			If two unitary representations are isomorphic, there exists a unitary $U$ such that $R_1(g)=UR_2(g)U^\dagger$.
		\end{thm}
		A proof can be found, for instance, in Ref.~\cite{Simon1996}, Theorem II.1.2. It is straightforward to show that the direct sum of two representations is again a representation. In turn, one may study how a representation decomposes into the direct sum of other representations. To this end, let us introduce the notion of an invariant subspace. A subspace $W$ of $V$ is called invariant if for all $g\in G$ and $w\in W$ it holds that $R(g)w\in W$. A representation restricted to an invariant subspace again gives a valid representation of the group, since this projection is a group homomorphism. Any representation has the trivial subspaces $W=V$ and $W={0}$. A representation is called irreducible (irrep) if it has only those trivial invariant subspaces. For finite groups, every irreducible representation is finite-dimensional. Moreover, for a unitary representation of a finite group, if $H$ is an invariant subspace, so is its orthogonal complement. One finds the following theorem.
		\begin{thm}\label{thm_decompositions}
			Any representation of a finite group is isomorphic to the direct sum of irreducible representations.
		\end{thm}
		Therefore, the irreps of a finite group provide complete information over all representations of that group. The number of inequivalent irreps of a finite group is determined by the conjugacy classes of that group.
		\begin{thm}\label{thm_conjug}
			The number of different conjugacy classes of a finite group is equal to the number of inequivalent irreducible representations of the group.
		\end{thm}
		Let us mention here that even though the number is the same, usually there is no natural bijection between the irreps of a group and its conjugacy classes. This is different for the symmetric group, as we discuss in Sec.~\ref{sec_irreps_sym}. Theorem~\ref{thm_conjug} motivates to study so called class functions. Those are functions from a group $G$ to the complex numbers, that have the same value for all elements of a conjugacy class. Note that the class functions of a group form a vector space on which an inner product can be defined via
		\begin{equation}
			\braket{f_{1}}{f_{2}}=\frac{1}{|G|}\sum_{g\in G} f_1(g^{-1})f_2(g),\label{eq_inner_product}
		\end{equation}
		where $f_{1,2}$ are class functions. An important class function is the character $\chi_R$ of a representation $R$. It is defined as
		\begin{equation}
			\chi_R(g)=\tr[R(g)]\ \ \forall g\in G.
		\end{equation}
		For any representation, the value of the character for the identity element $e$ is equal to the dimension of the representation. It can be shown that the characters of the irreps of a finite group form a basis for the vector space of class functions and that they are orthonormal with respect to the inner product in Eq.~\eqref{eq_inner_product}. Using that any representation of a finite group is equivalent to a unitary representation, we have that for $f_1,f_2$ characters, the inner product above can be rewritten as $\braket{f_{1}}{f_{2}}=\frac{1}{|G|}\sum_{g\in G} f_1(g)^* f_2(g)$. Additionally, characters are additive under direct sums of representations, i.e., $\chi_{R_1\oplus R_2}(g)=\chi_{R_1}(g)+\chi_{R_1}(g)$ for all $g\in G$. Combing this with Theorem~\ref{thm_decompositions}, it follows that the character of any representation is equal to the sum of the characters of the irreps it contains.

		Before we present a particular example of a representation, let us recall an important lemma from representation theory, namely Schur's lemma.
		\begin{lem}[Schur]\label{lem_schur}
			Let $D_1:G\rightarrow V_1$ and $D_2:G\rightarrow V_2$ be two irreducible representations and let $M:V_1\rightarrow V_2$ be a linear operator such that $M D_1(g)=D_2(g)M$ for all $g\in G$. Then $M=0$ or $M$ is an isomorphism. If additionally $D_1=D_2$, then there exists $\lambda \in\mathds{C}$ such that $M=\lambda \mathds{1}$.
		\end{lem}
		Thus, if a representation $D$ decomposes into the irreps $\{D_i\}$ with respective multiplicities $\{m_i\}$, i.e., 
		\begin{equation}
			D=\bigoplus_{i}\bigoplus_{j=1}^{m_i} D_i\simeq \bigoplus_{i}\mathds{1}_{m_i} \otimes D_i,\label{eq_decomp_irr}
		\end{equation}
		then any matrix $M\in M_{\dim(V)\times \dim(V)}(\mathds{C})$ that commutes with the representation of any group element is necessarily of the form
		\begin{align}
			M&=\bigoplus_{i}\begin{pmatrix}
				\lambda_{1,1}^i\mathds{1}_{\dim(D_i)}&\ldots&\lambda_{1,m_i}^i\mathds{1}_{\dim(D_i)}\\
				\vdots &&\vdots\\
				\lambda_{m_i,1}^i\mathds{1}_{\dim(D_i)}&\ldots&\lambda_{m_i,m_i}^i\mathds{1}_{\dim(D_i)}
			\end{pmatrix}\\
			&\simeq \bigoplus_{i} \begin{pmatrix}
				\lambda_{1,1}^i&\ldots&\lambda_{1,m_i}^i\\
				\vdots &&\vdots\\
				\lambda_{m_i,1}^i&\ldots&\lambda_{m_i,m_i}^i
			\end{pmatrix}\otimes \mathds{1}_{\dim(D_i)},\label{eq_decomposition_com}
		\end{align}
		where $\lambda_{k,l}^i\in\mathds{C}$ for all $i,k,l$.

		Lastly, let us introduce the so called regular representation. Before introducing the regular representation, let us first define the so called group algebra. Let $\mathds{C}(G)$ be the $\mathds{C}$ vector space that consists of all finite linear combinations of group elements with complex coefficients. Then, $\mathds{C}(G)$ forms an algebra, the so called group algebra of $G$. The group algebra (understood as a vector space) is the carrier space for the regular representation $R_{\rm reg}$. A linear invertible map $R_{\rm reg}(g)$, $g\in G$ acts on $\mathds{C}(G)$ via the group multiplication
		\begin{equation}
			R_{\rm reg}(g)\sum_{h\in G}c_h h=\sum_{h\in G}c_h (gh),
		\end{equation}
        where $c_h\in\mathbb{C}$ are some coefficients. Therefore, we have that $R_{\rm reg}(g)$ is given by $(R_{\rm reg}(g))_{k,l}=1$ if $g_k g_l^{-1}=g$ and $(R_{\rm reg}(g))_{k,l}=0$ else.
		
		From the above definition it follows that the character of the regular representation satisfies $\chi_{\rm reg}(e)=|G|$ and $\chi_{\rm reg}(g)=0$ for any $g\neq e$. Computing the inner product of this character with the character of any irrep, one finds that the regular representation contains any irrep, $R_i$, of the group exactly $d_i$ times, where $d_i$ denotes the dimension of $R_i$ for all $i$.

		\subsection{The symmetric group $S_N$}\label{app:symmetric_group}
		This appendix provides a brief introduction into nomenclature related to the symmetric group. A detailed introduction can be found in Ref.~\cite{fulton1991representation}. The symmetric group $S_N$ is the group formed by all permutations of $N$ distinct elements that we label by $1,\ldots, N$. Its order is $|S_N|=N!$. Every permutation can be written as a sequence of transpositions that act on $1,\ldots,N$ by exchanging two elements. The function ${\rm sgn}:S_N \rightarrow \{\pm 1\}$, called the sign of a permutation, is defined via
		\begin{equation}
			{\rm sgn}(\pi)=(-1)^{t},
		\end{equation}
		where $t$ is the number of transpositions required to generate $\pi$. Note that the parity of this number is well defined. We will see in the next section, that the sign function is a representation of the symmetric group $S_N$. The conjugacy classes of permutations in $S_N$ can be labeled by the integer partitions of $N$. An integer partition of $N$ is a sequence of positive integers that sum up to $N$. To label a partition we will in the following use a vector ${\bm \lambda} = (\lambda_1,\ldots,\lambda_n)$, where $n$ is the number of integers in the partition, $\lambda_i\in\mathbb{N}$ for all $i$ are the positive numbers such that $\sum_i \lambda_i=N$ and $\lambda_i\geq \lambda_{i+1}$ for all $i$.

		\subsection{Irreducible representations of the symmetric group}\label{sec_irreps_sym}
		The number of different conjugacy classes of a finite group is equal to the number of inequivalent irreps of that group, see Theorem~\ref{thm_conjug}. For $S_N$, the number of conjugacy classes is equal to the number of integer partitions of the number $N$, which is given by the so called partition function. A Young diagram is a pictorial way to represent such a partition.
		\begin{definition}
			The Young diagram that corresponds to the partition ${\bm \lambda}=(\lambda_1,\ldots,\lambda_n)$ of $N$ is a left-justified arrangement of $N$ boxes such that row $i$ has $\lambda_i$ boxes.
		\end{definition}
		If one assigns the numbers $1$ to $N$ to the boxes of a Young diagram, the resulting object is called a Young tableau $t$. A standard Young tableau is a Young tableau for which the numbers increase in each row from left to right and each column from top to bottom. Note that in general there is more than one standard Young tableau for a given Young diagram. As we will see below, different Young tableaus associated to the same Young diagram correspond to equivalent representations. For a Young tableau $t$, let us define the two subgroups $P_{t},Q_{t}\subseteq S_N$ formed by those permutations that do not change the numbers in each row/column of the tableau (but possibly permute them). Using those subgroups one can define the following three elements of the group algebra
		\begin{align}\label{eq_young_sym}
			a_{t}=\sum_{g\in P_{t}}g,\quad b_{t}=\sum_{g\in Q_{t}}\text{sgn}(g)g,\quad \mbox{ and }Y_{t}=a_{t}\cdot b_{t}.
		\end{align}
		Particularly important in the following is the element $Y_t$, the so called Young symmetrizer. In fact, we have the following theorem (see for instance Ref.~\cite{fulton1991representation} for a proof).
		\begin{thm}\label{thm_irreps_sym}
			The Young symmetrizer $Y_t$ is proportional to an idempotent element of the group algebra. Restricting the regular representation of $S_N$ to the subspace $\{Y_t a|a\in\mathds{C}(S_N)\}$, gives an irrep of $S_N$. Any irrep of $S_N$ can be obtained in this way from a unique Young tableau.
		\end{thm}
		The irreps obtained from two different Young tableaus associated to the same Young diagram are equivalent, see, for instance, Ref.~\cite{tung1985group} for a proof. The dimension of the irrep associated to a Young diagram is given by the number of standard Young tableaus that can be associated to the corresponding Young diagram. It can be computed by the hook length formula, which can be found for instance in Ref.~\cite{fulton1991representation}. As there are only two Young diagrams, $(N)$ and $(1,\ldots, 1)$, with only one associated standard Young tableau, there are only two inequivalent one dimensional representations of $S_N$. These are the trivial representation $D(\pi)=1$ and the sign representations $D(\pi)=\sgn(\pi)$ for all $\pi\in S_N$~\cite{fulton1991representation}. We present here the version of Ref.~\cite{i2016characterizing}.

		\subsection{Representation on tensor product spaces and Schur-Weyl duality \label{app_schurweyl}}
		We consider the unitary representation $D_N: S_N\rightarrow Gl(\mathcal{H})$, where $\mathcal{H}=(\mathbb{C}^M)^{\otimes N}$. Here permutations act by permuting tensor factors. It is defined by
		\begin{equation}\label{eq_rep_sn_tensor}
			D_N(\pi) \bigotimes_{i=1}^N\ket{\psi_i}=\bigotimes_{i=1}^N\ket*{\psi_{\pi(i)}},
		\end{equation}
		where $\bigotimes_{i=1}^N\ket*{\psi_{\pi(i)}}\in (\mathbb{C}^M)^{\otimes N}$. Even though the representation also depends on the local dimension $M$, we do not explicitly denote it as long as it does not lead to any ambiguity. In the following we will also omit the symbol $D_N$ if it is clear that we consider this particular representation to avoid notational overhead.
		
		Via the representation $D_N$, the Young symmetrizer corresponding to the Young diagram $(N)$ is mapped (up to normalization) to the projector onto the symmetric subspace
		\begin{equation}
			\PS{N}_S=\frac{1}{N!}\sum_{\pi\in S_N} D_N(\pi),
		\end{equation}
		and the Young symmetrizer corresponding to the Young diagram $(1,\ldots,1)$ is mapped (up to normalization) to the projector onto the antisymmetric subspace
		\begin{equation}
			\PS{N}_{A}=\frac{1}{N!}\sum_{\pi\in S_N}\sgn(\pi) D_N(\pi).
		\end{equation}
		The dimension of the symmetric subspace can be obtained from counting the number of linearly independent elements that can be obtained from applying $\PS{N}_S$ to the states of the computational basis (see for instance Ref.~\cite{Schwemmer2015} for a more detailed explanation). One finds that the dimension is given by
		\begin{equation}\label{eq_sim_sym}
			\dim(\mathcal{H}_{\rm sym})={N+M-1 \choose N}.
		\end{equation}
		The symmetric subspace $\mathcal{H}_{\rm sym}$ is the carrier space of the trivial representations contained in the decomposition of $D_N$ into irreps of $S_N$. That is, in the decomposition of $D_N$ into irreps, the one--dimensional trivial representation occurs $\dim(\mathcal{H}_{\rm sym})$ times and the corresponding vector space is the totally symmetric subspace. Using an analogous reasoning as above for $\PS{N}_{A}$ yields that the dimension of the antisymmetric subspace $\mathcal{H}_{\rm asym}$ is
		\begin{equation}\label{eq_dim_antisym}
			\dim(\mathcal{H}_{\rm asym})={M \choose N}
		\end{equation}
		if $N\leq M$ and $0$ otherwise. Similar to the symmetric subspace, the antisymmetric subspace is the carrier space for the alternating representations contained in the decomposition of $D_N$ into irreps of $S_N$.
		
		Next, let us study the decomposition of $D_N$ into irreps. One can show that there is a bijection between the irreps of the unitary group $U(M)$ and the symmetric group. This connection is captured by the so-called Schur-Weyl duality. First, let us introduce a representation $\tilde{D}_N:U(M)\rightarrow GL((\mathbb{C}^M)^{\otimes N})$ of the unitary group $U(M)$ defined by
		\begin{equation}\label{eq:rep_utn}
			\tilde{D}_N(U)=U^{\otimes N}.
		\end{equation}
		Clearly we have $\tilde{D}_N(U)D_N(\pi)=D_N(\pi)\tilde{D}_N(U)$ for all $\pi\in S_N$ and all $U\in U(M)$. Decomposing $D_N$ into irreps as in Eq.~\eqref{eq_decomp_irr} implies that $\tilde{D}_N$ is of the form as in Eq.~\eqref{eq_decomposition_com}. In this form, it is still possible that a subspace that carries all copies of a particular irrep of $S_N$, also carries multiple different irreps of $U(M)$. The Schur-Weyl duality states that any irrep of $S_N$ is paired with exactly one irrep of $U(M)$.
		
		\begin{thm}[Schur-Weyl duality]\label{thm_schur_weyl}
			Let $\tilde{D}_N$ and $D_N$ be the representation of the unitary group and the symmetric group on $(\bbC^M)^{\otimes N}$ defined by Eq.~\eqref{eq_rep_sn_tensor} and Eq.~\eqref{eq:rep_utn}, respectively. The space $(\bbC^M)^{\otimes N}$ decomposes as
			\begin{align}
				(\bbC^M)^{\otimes N}\simeq\bigoplus_{\lambda\vdash_M N}\HH_{\bm{\lambda}}\otimes \KK_{\bm{\lambda}}
			\end{align}
			such that $\tilde{D}_N$ acts as
			\begin{equation}
				\tilde{D}_N(U)\simeq \bigoplus_{\lambda\vdash_M N}  \tilde{I}_{\bm{\lambda}}(U)\otimes \mathds{1}_{\KK_{\bm{\lambda}}},
			\end{equation}
			and $D_N$ acts as
			\begin{equation}
				D_N(\pi)\simeq \bigoplus_{\lambda\vdash_M N} \mathds{1}_{\HH_{\bm{\lambda}}}\otimes I_{\bm{\lambda}}(\pi).
			\end{equation}
			Here, $\tilde{I}_{\bm{\lambda}}$ for all $\lambda$ are the irreps of $U(M)$, $I_{\bm{\lambda}}$ for all $\lambda$ are the irreps of $S_N$ and $\lambda\vdash_M N$ labels all Young diagrams with $N$ boxes and at most $M$ rows.
		\end{thm}
		The Schur-Weyl duality, together with Lemma~\ref{lem_schur} (Schur's lemma), implies that every permutation invariant state $\varrho\in\text{Inv}(\mathcal{H})$ must necessarily be of the form
		\begin{equation}\label{eq:SWdec}
			\varrho=\bigoplus_{\lambda\vdash_M N} p_{\bm{\lambda}} \varrho_{\bm{\lambda}}\otimes \frac{\id_{\KK_{\bm{\lambda}}}}{d_{\bm{\lambda}}},
		\end{equation}
		where $d_{\bm{\lambda}}$ is given by the hook length formula (see for instance Ref.~\cite{fulton1991representation}) and $\varrho_{\bm\lambda}$ is a state associated to every irrep. A proof for qubits can be found, for instance, in Proposition 5.1 in Ref.~\cite{Moroder2012}.
		
		Finally, let us discuss a concrete example for the decomposition in Eq.~\eqref{eq:SWdec} and the Schur-Weyl duality. We consider $3$ particles in $3$ modes and for simplicity we restrict ourselves to the subspace $\text{span}\{\pi\ket{123}|\pi\in S_3\}$ of the full Hilbert space. That is, we consider only states with a fixed mode occupation. The subspace $\text{span}\{\pi\ket{123}|\pi\in S_3\}$ decomposes into $4$ invariant subspaces of $S_N$. These are the symmetric subspace spanned by
		\begin{equation}
			\ket{\psi_1}=\ket{123}+\ket{132}+\ket{213}+\ket{231}+\ket{312}+\ket{321},
		\end{equation}
		which carries the trivial representation of $S_3$, the antisymmetric subspace spanned by
		\begin{equation}
			\ket{\psi_2}=\ket{123}-\ket{132}-\ket{213}+\ket{231}+\ket{312}-\ket{321},
		\end{equation}
		which carries the alternating representation of $S_3$, and two two-dimensional subspaces that both carry the two-dimensional irrep of $S_3$. These subspaces are spanned by
		\begin{align}
			\ket{\psi_3}&=\ket{123}+\ket{213}-\ket{231}-\ket{321}\\
			\ket{\psi_4}&=-\ket{123}-\ket{132}+\ket{312}+\ket{321}\\
		\end{align}
		and
		\begin{align}
			\ket{\psi_5}&=\ket{123}-\ket{132}-\ket{312}+\ket{321}\\
			\ket{\psi_6}&=\ket{123}-\ket{213}-\ket{231}+\ket{321},
		\end{align}
		respectively. We presented here, for simplicity, a nonorthogonal basis of $\text{span}\{\pi\ket{123}|\pi\in S_3\}$. Let $\ket{\tilde{\psi}_i}$ for $i=3,4,5,6$ denote an orthonormal basis, such that the representations on both subspaces defined above have the same matrix form. If $\varrho$ has support only on the subspace $\text{span}\{\pi\ket{123}|\pi\in S_3\}$ and is symmetric, i.e., $\varrho\in \text{Inv}(H)$, then Schur's lemma together with the above decomposition implies that $\varrho$ is necessarily of the form 
		\begin{align}
			\begin{split}
				\varrho&=\varrho_{(3)}\ketbra{\psi_1}+\varrho_{(1,1,1)}\ketbra{\psi_2}+(\varrho_{(2,1)})_{1,1}\ketbra{\tilde{\psi}_3}+(\varrho_{(2,1)})_{1,1}\ketbra{\tilde{\psi}_4}+(\varrho_{(2,1)})_{2,2}\ketbra{\tilde{\psi}_5}+\\&\ \ \ \ (\varrho_{(2,1)})_{2,2}\ketbra{\tilde{\psi}_6}
				+(\varrho_{(2,1)})_{1,2}\ketbra{\tilde{\psi}_3}{\tilde{\psi}_5}+(\varrho_{(2,1)})_{1,2}\ketbra{\tilde{\psi}_4}{\tilde{\psi}_6}+(\varrho_{(2,1)})_{2,1}\ketbra{\tilde{\psi}_5}{\tilde{\psi}_3}+(\varrho_{(2,1)})_{2,1}\ketbra{\tilde{\psi}_6}{\tilde{\psi}_4},
			\end{split}
			\\
			&=\begin{pmatrix}
				\varrho_{(3)}&0&0&0&0&0\\
				0&\varrho_{(1,1,1)}&0&0&0&0\\
				0&0&(\varrho_{(2,1)})_{1,1}&0&(\varrho_{(2,1)})_{1,2}&0\\
				0&0&0&(\varrho_{(2,1)})_{1,1}&0&(\varrho_{(2,1)})_{1,2}\\
				0&0&(\varrho_{(2,1)})_{2,1}&0&(\varrho_{(2,1)})_{2,2}&0\\
				0&0&0&(\varrho_{(2,1)})_{2,1}&0&(\varrho_{(2,1)})_{2,2}\\
			\end{pmatrix}
			\label{eq:eqloc1}
		\end{align}
		where $(\varrho_{(2,1)})_{2,1}=(\varrho_{(2,1)})_{1,2}^*$. This is precisely the form specified in Eq.~\eqref{eq:SWdec}. Note that in case the ONB $\{\ket{\tilde{\psi}_i}\}$ is chosen such that the representations on both two-dimensional subspaces do not have the same matrix form, then the sub-block $((\varrho_{3,5},\varrho_{3,6}),(\varrho_{4,5},\varrho_{4,6}))$ (and its conjugate) of $\varrho$ is no longer proportional to the identity. Instead it is given by the invertible matrix that relates the equivalent representations on both subspaces. Note that the state $\varrho$ in Eq.~\eqref{eq:eqloc1} describes perfectly indistinguishable particles according to our definition iff $\varrho_{(3)}=1$ and all other coefficients vanish. Moreover, $\varrho$ describes perfectly distinguishable particles according to our definition iff $(\varrho_{(2,1)})_{2,1}=0$ and $\varrho_{(3)}=\varrho_{(1,1,1)}=(\varrho_{(2,1)})_{1,1}=(\varrho_{(2,1)})_{2,2}=1/6$.
		
		\section{Properties of symmetric purifications}\label{app:purifications}
		
		In this section we are concerned with purifications of symmetric states, i.e., with purifications of states $\varrho\in {\rm Inv}(\mathcal{H})$. It is shown in Ref.~\cite{Renner2005} that for such states there always exists a purification $\ket{\psi}$ such that $\pi\otimes \tilde{\pi}\ket{\psi}=\ket{\psi}$ for all $\pi\in S_N$. Here, $\tilde{\pi}$ is the representation of the symmetric group that acts by permutation of the tensor factors on the purifying Hilbert space. In the following, we characterize for completeness all purifications of this type. In particular, we show that any $\ket{\psi}$ with the above properties is of the form $\ket{\psi}=\mathds{1}\otimes V \sqrt{\varrho}\ket{\varphi^+}$, where $V$ is an isometry $V$ between Hilbert spaces of $N$ particles such that $\tilde{\pi}V\vert_{\text{range}(\varrho^T)}=V\pi\vert_{\text{range}(\varrho^T)}$ for all $\pi\in S_N$ and $\ket{\varphi^+}$ is the Bell state.
		
		Let us consider an arbitrary purification $\ket{\psi}$ of the state $\varrho \in {\rm Inv}(\mathcal{H})$. It is straightforward to verify that any purification of $\varrho$ can be written as $\ket{\psi} = \id\otimes V\sqrt{\varrho}^T\ket*{\varphi^+}$, where $V$ is an isometry with $V^\dagger V=\id$. We restrict ourselves to isometries $V$ that map Hilbert spaces of $N$ particles to Hilbert spaces of $N$ particles. We are interested in purifications that satisfy $\pi\otimes\tilde{\pi}\ket{\psi}=\ket{\psi}$ for all $\pi\in S_N$. Using the symmetry of the state $\ket{\varphi^+}$, the condition $\pi\otimes\tilde{\pi}\ket{\psi}=\ket{\psi}$ holds if and only if 
		
		\begin{equation}
			\id\otimes \tilde{\pi}V\sqrt{\varrho}^T\pi^T\ket*{\varphi^+}=\id\otimes V\sqrt{\varrho}^T\ket*{\varphi^+}.\label{eq:IxStuff}
		\end{equation}
		
		It is straightforward to verify that if $\varrho$ is permutation invariant, then so is $\sqrt{\varrho}$ (see e.g. Ref.~\cite{Renner2005}). Thus, we have that $\sqrt{\varrho}^T=(\pi \sqrt{\varrho}\pi^\dagger)^T=\pi^*\sqrt{\varrho}^T\pi^T=\pi \sqrt{\varrho}^T\pi^\dagger$, where in the last equality we have used the fact that for the representation of $S_N$ in the tensor product space $(\mathbb{C}^M)^{\otimes N}$ all $\pi$ are real. Using this observation, we find that Eq.~\eqref{eq:IxStuff} holds if and only if $\tilde{\pi}V\pi^\dagger\sqrt{\varrho}^T  = V\sqrt{\varrho}^T$ holds on the range of $\sqrt{\varrho}^T$. Because the range of $\sqrt{\varrho_{\rm ext}}^T$ and $\varrho_{\rm ext}^T$ coincide the latter equality holds if and only if
		
		\begin{equation}
			\tilde{\pi}V\vert_{\text{range}(\varrho^T)}=V\pi\vert_{\text{range}(\varrho^T)}.\label{eq:final}
		\end{equation}
		
		Thus, we have shown that a purification $\ket{\psi} = \id\otimes V\sqrt{\varrho}^T\ket*{\varphi^+}$ of $\varrho$ satisfies $\pi\otimes \tilde{\pi}\ket{\psi}=\ket{\psi}$ for all $\pi\in S_N$ if and only if Eq.~\eqref{eq:final} holds for all $\pi\in S_N$.
		
		As already mentioned in the main text, the purification is not unique, and, e.g., the states $\ket{11}_{\rm ext}\otimes (\ket{12}+\ket{21})_{\rm int}$ and $\ket{11}_{\rm ext}\otimes \ket{11}_{\rm int}$ lead to the same $\varrho_{\rm ext}$. This also implies that interference experiments (accessing only external dofs) in general do not depend on whether the particles are individually addressable via their internal degrees of freedom, i.e., the total state (external and internal dofs) can be written as a superposition of states for which all internal states are orthogonal states. To make this more explicit note that for the first pair of photons a single photon can be addressed through a controlled operation with respect to the internal dofs while for the second pair of this is not possible.  This example also generalizes to more than two particles. If the reduced external state is bosonic, then also the reduced internal state has to be bosonic (due to the symmetry of the global state). Thus, for example, such a reduced external state is compatible with a reduced internal state of the form $\ket{k}^{\otimes N}$, where all particles occupy the same internal state. In case the dimension of the internal Hilbert space $\mathcal{H}_{\rm in}$ is larger or equal to the number of particles, then it is also compatible with the internal state $1/N!\sum_\pi \pi \ket{1,\ldots, N}$, where all particles are individually addressable.

		\section{Particle number measurement in first quantization}\label{app:red_to_qudits}
		
		In this section we show that particle number measurements on all output modes, which correspond to measurements of all the number operators $M_i=\sum_j a^\dagger_{ij} a_{ij}$, can be described by the operators
		\begin{equation}
			E(\vec{m})\otimes \mathds{1}=\sum_{\pi\in W(\vec{m})}\pi\ketbra{\vec{l}(\vec{m})}\pi^\dagger\otimes\mathds{1}
		\end{equation}
		in first quantization, where $\vec{m}$ are all the possible mode occupations, and $W(\vec{m})$ contains all permutations acting non-trivially on the mode assignment vector $\vec{l}(\vec{m})$. To this end, we first translate the description of the measurement in second quantization to first quantization and then show that the obtained measurement operator has the same action as $E(\vec{m})\otimes \mathds{1}$ on the Fock space.
		
		It is straightforward to see that the common eigenbasis of the operators $M_i=\sum_j a^\dagger_{ij} a_{ij}$ is given by the Fock states
		\begin{align}
			\ket*{\{m_i^j\}}&=\prod_{i,j}(a^\dagger_{i,j})^{m_i^j}\ket{0}\\
			&=\frac{1}{\sqrt{N!\prod_{k,l}m_k^l!}}\sum_{\pi\in S_N}\pi \ket{\vec{l}(\vec{m})}\otimes\tilde{\pi}\left[\bigotimes_{r=1}^P\bigotimes_{s=1}^M \ket{r}^{\otimes m_s^r}\right],
		\end{align}
		where $m_i^j\in \mathbb{N}$ is the number of particles in the external state $\ket{i}$, and the internal state $\ket{j}$, and $m_i=\sum_j m_i^j$ is the total number of particles in external state $\ket{i}$. Thus, if the occupation vector $\vec{m}$ with $\sum_i m_i=N$ is detected in the measurement of the operators $\{M_i\}$, the associated projector is given by 
		\begin{equation}
			\tilde{E}(\vec{m})=\sum_{\substack{m_{1}^1,\ldots ,m_{1}^P=0 \\ \sum_{j} m_1^j=m_1}}^{m_1}\ldots \sum_{\substack{m_{M}^1,\ldots ,m_{M}^P=0 \\ \sum_{j} m_M^j=m_M}}^{m_M}\ketbra*{\{m_{i}^j\}}.
		\end{equation}
		To establish the equality of $E(\vec{m})\otimes \mathds{1}$ and $\tilde{E}(\vec{m})$ restricted to the Fock space, we have to show that both operators have the same action on the basis vectors $\ket*{\{m_{i}^j\}}$, which is straightforward to verify.

		\section{Details to the proof of Theorem~\ref{thm:bounds}}\label{app:bounds}
		Here we provide the details to the proof of Theorem~\ref{thm:bounds}. We start with the optimization problem in Eq.~\eqref{eq:sdp} in the main text, and use the fact that $\PS{2}$ acting on any two particles $i\neq j$ can be expanded as $\PS{2}=\frac{1}{2}(\id+\pi_{ij})$, where $\pi_{ij}$ is the transposition in $S_N$ that exchanges particles $i$ and $j$. Thus, instead of requiring that $\mathrm{tr}[\PS{2}\varrho]=p_2$, we can require that $\tr[\pi_{ij}\varrho]=2p_2-1=c$. Due to the permutation invariance of $\varrho$ this has to hold for all transpositions $\pi_{ij}$, and hence, summing over all $\pi_{ij}$ and dividing by the number of permutations one has transformed the last condition to $\tr[T\varrho]=c$, where $T=\frac{2}{N(N-1)}\sum_{i\neq j}\pi_{ij}$. To simplify the optimization problem in Eq.~\eqref{eq:sdp} we need to know the spectrum of $T$.
    
		From Ref.~\cite{Rothaus1966} one can infer that the spectrum of $T$ (up to multiplicities that are irrelevant for us and scaling factors (see below)) is given by
		\begin{equation}
			c_{\bm{\lambda}} = \frac{1}{2}\qty[\sum_j\lambda_j(\lambda_j-1) - \sum_j\lambda^T_j(\lambda^T_j-1)],
		\end{equation}
		where $\bm{\lambda}^T$ denotes the conjugate partition of $\bm{\lambda}$, which is the partition corresponding to the transposed Young diagram of $\bm{\lambda}$. Finally, taking into account the normalization of $T$, and using the relation derived in Eq.~(D-2) in Ref.~\cite{Diaconis1988} one arrives at the following eigenvalues
		\begin{equation}\label{eq:eigenvalues}
			t_{\bm{\lambda}} = \frac{1}{N(N-1)}\sum_j\qty[\lambda_j^2-(2j-1)\lambda_j].
		\end{equation}
		Decomposing $\varrho=\bigoplus_{\bm{\lambda}}p_{\bm{\lambda}}\varrho_{\bm{\lambda}}$ in a direct sum over the eigenspaces of $T$ we have arrived at the linear program
  \begin{equation}
		\begin{aligned}
			b= \min \Bigl\{ p_{(N)}: \sum_{\bm{\lambda}} p_{\bm{\lambda}} =1, \sum_{\bm{\lambda}} t_{\bm{\lambda}} p_{\bm{\lambda}} =c, p_{\bm{\lambda}}\geq 0\Bigr\},
		\end{aligned}   
	\end{equation}
 where we have used that the subspace corresponding to $t_{(N)}=+1$ is precisely the symmetric subspace, $(N)\equiv(N,0,\dots,0)$, and $p_{(N)}=p_N$. The linear program can be solved using Lagrangian multipliers. Choosing $p_{\bm{\lambda}} = q_{\bm{\lambda}}^2$ to get rid of the positivity constraint we obtain the Lagrange function $L(\qty{q_{\bm{\lambda}}},\mu,\nu)=q_{(N)}^2 - \mu (\sum_{\bm{\lambda}} q_{\bm{\lambda}}^2 -1)-\nu (\sum_{\bm{\lambda}} t_{\bm{\lambda}} q^2_{\bm{\lambda}} -c)$ and we wish to solve $\nabla L=0$. One finds the following set of equations
		\begin{align}
			q_{(N)}(1-\mu -\nu )&=0,\\
			q_{\bm{\lambda}}(\mu+\nu t_{\bm{\lambda}})&=0,\ \forall\lambda\neq (N)\label{eq:mult}\\
			\sum_{\bm{\lambda}} q_{\bm{\lambda}}^2 &=1,\\
			\sum_{\bm{\lambda}} t_{\bm{\lambda}} q^2_{\bm{\lambda}} &=c,
		\end{align}
		where $(N)=(N,0,\dots,0)$. The set of solutions can be divided into two cases.
		
		\emph{(Case 1)} In Eq.~\eqref{eq:mult} two or more of the $q_{\bm{\lambda}}$ are non-zero. Then, due to the fact that the $t_{\bm{\lambda}}$ are all pairwise distinct, it follows that $\mu=\nu=0$, and hence $q_{(N)}=0$ and $p_{(N)}=0$. Moreover, from the last equation we find that $c\leq t_{(N-1,1)}$, which is the second largest eigenvalue of $T$. Using Eq.~\eqref{eq:eigenvalues} we find $t_{(N-1,1)} = (N-3)/(N-1)$.
		
		\emph{(Case 2)} In Eq.~\eqref{eq:mult} exactly one of the $q_{\bm{\lambda}}$ is non-zero. But then we have that $p_{(N)}+p_{\bm{\lambda}}=1$ and $p_{(N)}+t_{\bm{\lambda}}p_{\bm{\lambda}}=c$, hence $p_{(N)}=(c-t_{\bm{\lambda}})/(1-t_{\bm{\lambda}})$, which is minimized for $t_{\bm{\lambda}}=t_{(N-1,1)}$.
		
		Moreover, we can replace the minimization in Eq.~\eqref{eq:sdp} by a maximization and ask what is the smallest upper bound $p_N\leq B$ for a given value of $p_2$. Since the Lagrange equations are the same we are just left with maximizing $p_{(N)}=(c-t_{\bm{\lambda}})/(1-t_{\bm{\lambda}})$, which is maximized by the smallest $t_{\bm{\lambda}}=-1$, and therefore $B=p_2$.

        Tightness of the bounds follows from the fact that the state $\alpha \ketbra{D_{N}}+(1-\alpha)\ketbra{AS_{N}}$, where $\ket{D_{N}}\propto \sum_\pi \pi\ket{1\dots N}$ is a Dicke state, and $\ket{AS_{N}}\propto \sum_\pi \text{sign}(\pi)\pi\ket{1\dots N}$ is the totally anti-symmetric state, saturates the upper bound for $\alpha = p_2$, and the non-zero lower bound for $\alpha = p_2(N-1)-(N-2)$. Since this state has a fixed mode occupation, our bounds remain tight even if one restricts to a fixed mode occupation.

		\section{Proof of Theorem~\ref{thm:distinguishable}}\label{app:distinguishability}

		We prove Theorem~\ref{thm:distinguishable} in two steps. First, we show that any eigenspace (to a non--vanishing eigenvalue) of a state describing perfectly distinguishable particles, has the property that if one restricts the unitary representation, $D_N$ (see Eq.~\eqref{eq_rep_sn_tensor}), to this subspace one obtains an integer number of copies of the regular representation. Then, we show that projectors onto subspaces with this property are, up to a permutationally invariant unitary, direct sums of projectors onto computational basis states and their permutations. 
		
		Let us start by writing the perfectly distinguishable state $\varrho_{\rm ext}$ in its spectral decomposition $\varrho_{\rm ext} = \sum_{k}q_k P_k$. Due to our definition of perfect distinguishability we have that for any projector $P_k$ on the eigenspace $V_k$ it holds that
		\begin{eqnarray}\label{eq:character_dist}
			\tr[\pi P_k] = \begin{cases}
				\dim(V_k) &\text{if } \pi=\text{id},\\
				0 \quad &\text{otherwise}.
			\end{cases}
		\end{eqnarray}
		If we consider the unitary representation $\pi$ of the permutation group on $(\mathbb{C}^M)^{\otimes N}$, and restrict it to the subspace $V_k\subset (\mathbb{C}^M)^{\otimes N}$ we obtain a unitary representation $\pi_k=\pi\vert_{V_k}=P_k\pi P_k$ of the permutation group, since $V_k$ is an invariant subspace. For its character one finds $\chi[\pi_k] = \tr[\pi_k]=\tr[\pi P_k]$, which is given in Eq.~\eqref{eq:character_dist}.
		
		We now show that any of these representations, $\pi_k$, is isomorphic to the direct sum of $i_k\in\mathbb{N}$  copies of the regular representation. We do so by showing that their characters coincide \footnote{Recall that the characters of inequivalent irreps are mutually orthogonal. Hence, the character of a representation defines uniquely its decomposition into irreps.}. The character of the regular representation is given by $\chi_{\rm reg}[\id_{\rm reg}] = |S_N|$, and $\chi_{\rm reg}[\pi_{\rm reg}] = 0$ otherwise. Given that the character of $\pi_k$ vanishes for any permutation but the identity, it remains to show that the dimension of the representation $\mathrm{dim}(V_k)=\chi[\id_k]$ is divisible by the order of the group. To show that, we consider the inner product of $\chi(\pi_k)$ and the character of the trivial representation, defined as $\pi_{\rm tr}=1$, $\chi_{\rm tr}$. We find 
		\begin{equation}
			\langle\chi(\pi_k),\chi_{\rm tr} \rangle = \frac{1}{|S_N|}\sum_{\pi\in S_N}\overline{\chi(\pi_k)}\chi_{\rm tr}(\pi_{\rm tr})=\frac{\chi[\id_k]}{|S_N|}=i_k.
		\end{equation}
		As the one--dimensional trivial representation is irreducible, $i_k$ is an integer. It is the number of times the trivial irrep is contained in the representation $\pi_k$. This shows that the character of $\pi_k$ is given by $\chi = i_k\cdot\chi_{\rm reg}$ and therefore, $\pi_k$ is isomorphic to the direct sum of $i_k$ copies of the regular representation.

		It remains to be shown that the above implies that $\varrho$ up to a permutationally invariant unitary is given by a direct sum of projectors onto computational basis states (with no two particles in the same state) and their permutations. To show this, we explicitly construct a unitary $U$ that does the transformation and is permutationally invariant. To this end, let us split up $(\bbC^M)^{\otimes N}$ into invariant subspaces of $D_N$ in two ways. From the above discussion it follows that one can decompose $(\bbC^M)^{\otimes N}$ into invariant subspaces as 
		\begin{equation}
			(\bbC^M)^{\otimes N}=(\bigoplus_k\bigoplus_{l=1}^{i_k} R_{k,l})\oplus \text{ker}(\varrho).\label{eq:dec1}
		\end{equation}
		Here, $R_{k,l}$ for all $l$ labels the subspaces in the eigenspace $V_k$ of $\varrho$, that carry the regular representation and $\text{ker}(\varrho)$ is the kernel of $\varrho$. To write down another decomposition of $(\bbC^M)^{\otimes N}$ into invariant subspaces, for each subspace $R_{k,l}$ let us choose a different computational basis states $\ket{\vec{b}(k,l)}\in (\bbC^M)^{\otimes N}$ with $b(k,l)_i< b(k,l)_{i+1}$ for all $i\in\{1,\ldots, M-1\}$. 
		The fact that there are sufficiently many different computational basis states of this form can be seen as follows: Each regular representation in $\text{range}(\varrho)$ carried by $R_{k,l}$ contains the sign representation exactly once. Hence, $\sum_k i_k$ is at most equal to the dimension of the antisymmetric subspace, which is spanned by antisymmetric combinations of the vectors $\ket{b(k,l)_i}$. Hence, $\sum_k i_k$ lower bounds the number of computational basis states of the desired form. Using these computational basis states we find
		\begin{equation}
			(\bbC^M)^{\otimes N}=(\bigoplus_k\bigoplus_{l=1}^{i_k}\text{span}\{D_N(\pi)\ket{\vec{b}(k,l)}\}_\pi)\oplus C,\label{eq:dec2}
		\end{equation}
		where $C$ is the orthogonal complement of $\bigoplus_k\bigoplus_{l=1}^{i_k}\text{span}\{D_N(\pi)\ket{\vec{b}(k,l)}\}_\pi$ in $(\bbC^M)^{\otimes N}$. It is straightforward to verify that the subspace $\text{span}\{D_N(\pi)\ket{\vec{b}(k,l)}\}_\pi$ for each $k,l$ is an invariant subspace of $D_N$ and carries the regular representation of $D_N$. As $D_N$ restricted to each of these subspaces is a unitary representation, also $C$, as the orthogonal complement of these subspaces, is an invariant subspace. Moreover, the representation $D_N$ restricted to the subspace $C$ ($D_N|_C$) is unitary. It follows from Eq.~\eqref{eq:dec1} and \eqref{eq:dec2} that the representation $D_N|_{C}$ is equivalent to the representation $D_N|_{\text{ker}(\varrho)}$. Thus, Theorem~\ref{thm_unitary} implies that there exists a unitary transformation $U: \text{ker}(\varrho)\rightarrow C $ that relates the subrepresentations $D_N|_{\text{ker}(\varrho)}$ and $D_N|_{C}$. By the same argument, such a unitary transformation also exists between the subrepresentations $D_N|_{R_{k,l}}$ and $D_N|_{\text{span}\{D_N(\pi)\ket{\vec{b}(k,l)}\}_\pi}$ for all $k,l$. Thus, we have
		\begin{align}
			D_N(\pi)&=(\bigoplus_k\bigoplus_{l=1}^{i_k} D_N|_{R_{k,l}})\oplus D_N|_{\text{ker}(\varrho)}\\
			&=(\bigoplus_k\bigoplus_{l=1}^{i_k} D_N|_{\text{span}\{D_N(\pi)\ket{\vec{b}(k,l)}\}_\pi})\oplus D_N|_{C}\\
			&=(\bigoplus_k\bigoplus_{l=1}^{i_k} U(k,l)D_N|_{R_{k,l}})U(k,l)\oplus U D_N|_{\text{ker}(\varrho)}U^\dagger\\
			&=V D_N(\pi) V^\dagger,
		\end{align}
		where $V=(\bigoplus_k\bigoplus_{l=1}^{i_k}U(k,l))\oplus U$. That is, the unitary $V$ is permutationally invariant and maps $D_N|_{R_{k,l}}$ to $D_N|_{\text{span}\{D_N(\pi)\ket{\vec{b}(k,l)}\}_\pi}$. As $\varrho$ acts as the identity on each of the subspaces $R_{k,l}$, $V\varrho V^\dagger$ is of the desired form.

		\section{Relation to other definitions of (in)distinguishability and to many-body interference}\label{app:relation}
		
		In this section we discuss the relation of our definitions of (in)distinguishability to other definitions in the literature. Moreover, we discuss the role of indistinguishability, as defined here, in the emergence of many-body interference in experiments consisting of linear interferometers and particle-number detection. To make the following discussion easier to follow let us recall the definitions of indistingusihability and distinguishability from the main text.

        \begin{definition}\label{def:(in)distinguishability}
            A state $\varrho\in \text{Inv}(\HH_{\text{ext}})$ describes indistinguishable particles if $P_{(N)}\varrho=\varrho$. The state $\varrho$ describes distinguishable particles if $\tr(D_N(\pi) P_k)=0$ for all $\pi\in S_N$ with $\pi\neq \mathds{1}$ and for all $P_k$, where $\{P_k\}$ are the projectors onto the eigenspaces of $\varrho$ to all non-zero eigenvalues.
        \end{definition}
  
		\subsection{Comparison to other definitions from the literature\label{app:comparison_literature}}
		
		First, let us mention that there are several definitions for what it means for particles to be perfectly (in)distinguishable. In the following be discuss two commonly used definitions and compare our definition with them. Note however, that there exist even more defintions (see for instance Ref.~\cite{Ou2006}) and we refer the reader to Ref~\cite{Dittel2019} for a comparison of further different definitions. Let us stress that the two commonly used definitions, which we discuss in the following, only apply to particular subsets of states.

		In Ref.~\cite{Shchesnovich2015}, perfect (in)distinguishability is defined for states with a fixed external mode occupation, i.e., states that are eigenstates of the particle number operator $M_j =\sum_k a_{jk}^\dagger a_{jk}$ for all $j$. To recall the definition presented in Ref.~\cite{Shchesnovich2015}, we consider states with a fixed occupation of the external modes $\vec{m}$. According to Ref.~\cite{Shchesnovich2015}, particles described by such states are perfectly indistinguishable if \begin{equation}
			\ket{\psi}\propto \sum_{\pi\in S_N} \pi\ket{\vec{l}(\vec{m})},
		\end{equation}
		where $\vec{l}(\vec{m})\in[M]^N$ is the mode assignment vector corresponding to the occupation vector $\vec{m}$. Thus, for instance, two particles described by the state $\ket{\psi}=1/\sqrt{2}(\ket{12}+\ket{21})$ are indistinguishable. Note that these states are typically called Dicke states~\cite{Dicke1954}. Moreover, in Ref.~\cite{Shchesnovich2015} particles are distinguishable, if $\varrho$ is diagonal in the bases generated by the external modes. It is straightforward to verify that Definition~\ref{def:(in)distinguishability} agrees with the definition of perfect indistinguishability of Ref.~\cite{Shchesnovich2015} when restricted to states with a fixed occupation of the external modes. However, Definition~\ref{def:(in)distinguishability} differs from the definition of perfect distinguishability of Ref.~\cite{Shchesnovich2015} when restricted to states with a fixed occupation of the external modes. According to Definition~\ref{def:(in)distinguishability} a state $\varrho$ with fixed occupation of the external modes $\vec{m}$ describe perfectly distinguishable particles if it is of the form
		\begin{equation}
			\varrho=\frac{1}{N!} \sum_{\pi\in S_N}\pi\ketbra{\vec{l}(\vec{m})}\pi^\dagger,
		\end{equation}
		with $m_i\le 1$ for all $i$. Thus, in contrast to Ref.~\cite{Shchesnovich2015} only diagonal states with at most one particle per mode describe perfectly distinguishable particles. We justify our definition of perfectly distingusihable particles further in Appendix~\ref{app:motivation}. For instance, the state $\varrho\propto\ketbra{112}+\ketbra{121}+\ketbra{211}$ describes distinguishable particles according to Ref.~\cite{Shchesnovich2015}, while it does not according to Definition~\ref{def:(in)distinguishability}.

		Our definition also stands partially in contrast to perfect distinguishability as defined in, e.g., Refs.~\cite{benatti2020entanglement}. There, particles are called perfectly distinguishable if they are described by states of the form $\sum_{m_1,\ldots m_N=1}^Nc(m_1,\ldots,m_N) a_{m_1,\psi_1}^\dagger\ldots a_{m_N,\psi_N}$, where $c(m_1,\ldots,m_N)\in\mathbb{C}$ and $\ket{\psi_1},\ldots,\ket{\psi_N}\in\HH_\text{int}$ are the internal states of the particles and $\braket{\psi_i}{\psi_j}=\delta_{ij}$. This definition of perfect distinguishability deviates from Definition~\ref{def:(in)distinguishability}. To illustrate this, let us consider the state $\ket{\psi}\propto(\ket{12}+\ket{21})\otimes (\ket{12}+\ket{21})$ as an example. Clearly, in this state one particle occupies the internal state $\ket{1}$ and the other particle the state $\ket{2}$. Thus, those particles are perfectly distinguishable according to the above definition. However, observe that $\ket{\psi}_{\rm ext}\propto\ket{12}+\ket{21}$ and therefore the particles are perfectly indistinguishable according to Definition~\ref{def:(in)distinguishability}. That is, whether particles occupy orthogonal internal states, i.e., whether particles are individually addressable with respect to the internal dofs, is in general not accessible by linear interferometers and particle-number measurements on the external modes.

		\subsection{The role of indistinguishability in the emergence of many-body interference}
		
		Let us mention that our definition of perfect (in)distinguishability is not related to whether a state can display many-body interference~\cite{dufour2020many,brunner2023many} or not. To see this, consider the states 
		\begin{equation}
			\varrho_{\rm ext}^a=\ketbra{1,\ldots, 1}
		\end{equation}
		and
		\begin{equation}
			\varrho_{\rm ext}^b=\frac{1}{N!}\sum_{\pi,\sigma\in S_N}\pi\ketbra{1,\ldots, N}\sigma^\dagger.
		\end{equation}
		For both states the particles are perfectly indistinguishable according to Definition~\ref{def:(in)distinguishability}. However, only $\varrho_{ex}^b$ allows for the observation of many-particle interference. Indeed, recall from Section~\ref{app:red_to_qudits} that the particle-number measurement with respect to the external modes is equivalent to a computational-basis measurement. As $\varrho_{ex}^a$ is separable and also the interferometer implements a local unitary, any measurement outcome can be explained with classically correlated particles. 
		
		We see from the above discussion that particle entanglement of $\varrho_{ex}$ is crucial for observing more than single-particle interference when the respective state is sent through a non-interacting interferometer and then the particle number is measured with respect to the external modes. Moreover, it can be shown~\cite{MatthiasThesis} that using non-interacting interferometers and destructive particle-number measurements with respect to some subset of external modes (i.e., with post-selection on some modes), only separable states can be generated in case the initial state was separable. Clearly, if non-local unitaries can be implemented through the interferometer, then this is no longer the case. The same also holds if the particle-number measurement is non-destructive. This can be seen with the help of the following example. Consider the separable state $\varrho_{\rm ext}=\ketbra{11}$. Applying a beam splitter, i.e., applying the unitary $U^{\otimes 2}$ with $U=1/\sqrt{2}((1,1),(1,-1))$, and post-selecting on detecting a single particle in each output mode yields the entangled state $\varrho_{\rm ext}'\propto(\ket{12}+\ket{21})(\bra{12}+\bra{21})$.

		\section{Motivation for definition of perfect distinguishability}\label{app:motivation}
		
		In the following we describe in more detail the motivation for our definition of perfectly distinguishable particles (see Definition~\ref{def:(in)distinguishability}). In particular, we show that our definition provides a meaningful negation of the definition of perfectly indistinguishable particles.
		
		To this end, let us briefly recall the definition of perfectly indistinguishable particles. We say (i) particles are perfectly indistinguishable if $\PS{N}\varrho=\varrho$ (see Definition~\ref{def:(in)distinguishability}). As we discussed in the main text the latter condition holds (ii) if and only if $\pi\varrho=\varrho$ for all $\pi\in S_N$. Expanding $\varrho$ in its eigendecomposition, $\varrho=\sum_k q_k P_k$, it is straightforward to show that (iii) particles are perfectly indistinguishable if and only if $\pi P_k=P_k$ for all $k$ and $\pi\in S_N$.
		
		States that describe perfectly distinguishable particles as the other extreme type of state besides those that describe perfectly indistinguishable particles should maximally violate one of these equivalent conditions. Clearly, our definition of perfectly distinguishable particles (Definition~\ref{def:(in)distinguishability}) negates condition (iii) in the strongest possible way. Let us discuss why the negation of the other two conditions leads to a definition of perfectly distinguishable states with undesirable properties. To negate condition (i) one could demand that a state $\varrho$ describes perfectly distinguishable particles if $P^{(N)}\varrho_{\rm ext}=0$, that is if $\varrho$ does not have support on the symmetric subspace. However, then states that are diagonal in the basis of external modes are not perfectly distinguishable, even though such states were classified as perfectly distinguishable in other works (see Section~\ref{app:comparison_literature}). There are two meaningful negations of the condition (ii). On the one hand one could demand that $\tr(\pi\varrho_{\rm ext})=0$ for all $\pi\in S_N$ with $\pi\neq \id$. Alternatively, one could negate the condition by asking that upon application of a permutation, the state $\varrho$ should become orthogonal to itself, i.e., $\tr(\varrho \pi\varrho)=\tr(\varrho^2 \pi)=0$ for all $\pi\in S_N$ with $\pi\neq\id$. However, each of the conditions can be satisfied by states with a fixed occupation of the external modes that are not perfectly distinguishable according to the definition of Ref.~\cite{Shchesnovich2015}. According to the latter reference the only perfectly distinguishable states for fixed occupation of the external modes are those that are diagonal in the basis generated by the external modes. Examples of states that satisfy $\tr(\pi\varrho_{\rm ext})=0$ or $\tr(\varrho \pi\varrho)=\tr(\varrho^2 \pi)=0$ for all $\pi\in S_N$ with $\pi\neq \mathds{1}$ and that have a fixed occupation of the external modes exist already for $N=3$ particles. Consider for instance the state with matrix representation

        \begin{equation}
            \varrho=\begin{pmatrix}
                1/6&1/9& -5/90& 0&0&-5/90\\
                1/9&1/6&0&-5/90&-5/90&0\\
                -5/90&0&1/6&-5/90&1/9&0\\
                0&-5/90&-5/90&1/6&0&1/9\\
                0&-5/90&1/9&0&1/6&-5/90\\
                -5/90&0&0&1/9&-5/90&1/6
            \end{pmatrix}
        \end{equation}
  
        with respect to the basis $\{\pi\ket{123}\}$. It is straightforward to verify that $\tr(\varrho \pi)=0$ for all $\pi\in S_3$ with $\pi\neq \mathds{1}$. Similarly, the state $\varrho'\propto\sqrt{\varrho}$ satisfies $\tr((\varrho')^2 \pi)=0$ for all $\pi\in S_3$ with $\pi\neq \mathds{1}$.

		As we already noted in the main text, the set of states that describe distinguishable particles according to our definition is not convex. The convex hull of this  set is contained in the convex set of states that satisfy $\mathrm{tr}[\varrho P^{(N)}]=1/N!$. However, it is currently unclear whether the convex hull of states that describe distinguishable particles and the set of states that satisfy $\mathrm{tr}[\varrho P^{(N)}]=1/N!$ agree. A visualization of the above discussion is presented in Fig.~\ref{fig:convex}.


\begin{thebibliography}{99}
			\makeatletter
			\providecommand \@ifxundefined [1]{%
				\@ifx{#1\undefined}
			}%
			\providecommand \@ifnum [1]{%
				\ifnum #1\expandafter \@firstoftwo
				\else \expandafter \@secondoftwo
				\fi
			}%
			\providecommand \@ifx [1]{%
				\ifx #1\expandafter \@firstoftwo
				\else \expandafter \@secondoftwo
				\fi
			}%
			\providecommand \natexlab [1]{#1}%
			\providecommand \enquote  [1]{``#1''}%
			\providecommand \bibnamefont  [1]{#1}%
			\providecommand \bibfnamefont [1]{#1}%
			\providecommand \citenamefont [1]{#1}%
			\providecommand \href@noop [0]{\@secondoftwo}%
			\providecommand \href [0]{\begingroup \@sanitize@url \@href}%
			\providecommand \@href[1]{\@@startlink{#1}\@@href}%
			\providecommand \@@href[1]{\endgroup#1\@@endlink}%
			\providecommand \@sanitize@url [0]{\catcode `\\12\catcode `\$12\catcode
				`\&12\catcode `\#12\catcode `\^12\catcode `\_12\catcode `\%12\relax}%
			\providecommand \@@startlink[1]{}%
			\providecommand \@@endlink[0]{}%
			\providecommand \url  [0]{\begingroup\@sanitize@url \@url }%
			\providecommand \@url [1]{\endgroup\@href {#1}{\urlprefix }}%
			\providecommand \urlprefix  [0]{URL }%
			\providecommand \Eprint [0]{\href }%
			\providecommand \doibase [0]{http://dx.doi.org/}%
			\providecommand \selectlanguage [0]{\@gobble}%
			\providecommand \bibinfo  [0]{\@secondoftwo}%
			\providecommand \bibfield  [0]{\@secondoftwo}%
			\providecommand \translation [1]{[#1]}%
			\providecommand \BibitemOpen [0]{}%
			\providecommand \bibitemStop [0]{}%
			\providecommand \bibitemNoStop [0]{.\EOS\space}%
			\providecommand \EOS [0]{\spacefactor3000\relax}%
			\providecommand \BibitemShut  [1]{\csname bibitem#1\endcsname}%
			\let\auto@bib@innerbib\@empty
			%</preamble>
			\bibitem [{\citenamefont {Hong}\ \emph {et~al.}(1987)\citenamefont {Hong},
				\citenamefont {Ou},\ and\ \citenamefont {Mandel}}]{Hong1987}%
			\BibitemOpen
			\bibfield  {author} {\bibinfo {author} {\bibfnamefont {C.~K.}\ \bibnamefont
					{Hong}}, \bibinfo {author} {\bibfnamefont {Z.~Y.}\ \bibnamefont {Ou}}, \ and\
				\bibinfo {author} {\bibfnamefont {L.}~\bibnamefont {Mandel}},\ }\href
			{\doibase 10.1103/PhysRevLett.59.2044} {\bibfield  {journal} {\bibinfo
					{journal} {Phys. Rev. Lett.}\ }\textbf {\bibinfo {volume} {59}},\ \bibinfo
				{pages} {2044} (\bibinfo {year} {1987})}\BibitemShut {NoStop}%
			\bibitem [{\citenamefont {Knill}\ \emph {et~al.}(2001)\citenamefont {Knill},
				\citenamefont {Laflamme},\ and\ \citenamefont {Milburn}}]{Knill2001}%
			\BibitemOpen
			\bibfield  {author} {\bibinfo {author} {\bibfnamefont {E.}~\bibnamefont
					{Knill}}, \bibinfo {author} {\bibfnamefont {R.}~\bibnamefont {Laflamme}}, \
				and\ \bibinfo {author} {\bibfnamefont {G.~J.}\ \bibnamefont {Milburn}},\
			}\href {\doibase 10.1038/35051009} {\bibfield  {journal} {\bibinfo  {journal}
					{Nature}\ }\textbf {\bibinfo {volume} {409}},\ \bibinfo {pages} {46}
				(\bibinfo {year} {2001})}\BibitemShut {NoStop}%
			\bibitem [{\citenamefont {Aaronson}\ and\ \citenamefont
				{Arkhipov}(2013)}]{Aaronson2013}%
			\BibitemOpen
			\bibfield  {author} {\bibinfo {author} {\bibfnamefont {S.}~\bibnamefont
					{Aaronson}}\ and\ \bibinfo {author} {\bibfnamefont {A.}~\bibnamefont
					{Arkhipov}},\ }\href {\doibase 10.4086/toc.2013.v009a004} {\bibfield
				{journal} {\bibinfo  {journal} {Theory Comput.}\ }\textbf {\bibinfo {volume}
					{9}},\ \bibinfo {pages} {143} (\bibinfo {year} {2013})}\BibitemShut {NoStop}%
			\bibitem [{\citenamefont
				{Shchesnovich}(2015{\natexlab{a}})}]{Shchesnovich2015}%
			\BibitemOpen
			\bibfield  {author} {\bibinfo {author} {\bibfnamefont {V.~S.}\ \bibnamefont
					{Shchesnovich}},\ }\href {\doibase 10.1103/PhysRevA.91.013844} {\bibfield
				{journal} {\bibinfo  {journal} {Phys. Rev. A}\ }\textbf {\bibinfo {volume}
					{91}},\ \bibinfo {pages} {013844} (\bibinfo {year}
				{2015}{\natexlab{a}})}\BibitemShut {NoStop}%
			\bibitem [{\citenamefont {Brunner}\ \emph {et~al.}(2022)\citenamefont
				{Brunner}, \citenamefont {Buchleitner},\ and\ \citenamefont
				{Dufour}}]{Brunner2022}%
			\BibitemOpen
			\bibfield  {author} {\bibinfo {author} {\bibfnamefont {E.}~\bibnamefont
					{Brunner}}, \bibinfo {author} {\bibfnamefont {A.}~\bibnamefont
					{Buchleitner}}, \ and\ \bibinfo {author} {\bibfnamefont {G.}~\bibnamefont
					{Dufour}},\ }\href {\doibase 10.1103/PhysRevResearch.4.043101} {\bibfield
				{journal} {\bibinfo  {journal} {Phys. Rev. Res.}\ }\textbf {\bibinfo {volume}
					{4}},\ \bibinfo {pages} {043101} (\bibinfo {year} {2022})}\BibitemShut
			{NoStop}%
			\bibitem [{\citenamefont {Brod}\ \emph {et~al.}(2019)\citenamefont {Brod},
				\citenamefont {Galv\~ao}, \citenamefont {Viggianiello}, \citenamefont
				{Flamini}, \citenamefont {Spagnolo},\ and\ \citenamefont
				{Sciarrino}}]{Brod2019}%
			\BibitemOpen
			\bibfield  {author} {\bibinfo {author} {\bibfnamefont {D.~J.}\ \bibnamefont
					{Brod}}, \bibinfo {author} {\bibfnamefont {E.~F.}\ \bibnamefont {Galv\~ao}},
				\bibinfo {author} {\bibfnamefont {N.}~\bibnamefont {Viggianiello}}, \bibinfo
				{author} {\bibfnamefont {F.}~\bibnamefont {Flamini}}, \bibinfo {author}
				{\bibfnamefont {N.}~\bibnamefont {Spagnolo}}, \ and\ \bibinfo {author}
				{\bibfnamefont {F.}~\bibnamefont {Sciarrino}},\ }\href {\doibase
				10.1103/PhysRevLett.122.063602} {\bibfield  {journal} {\bibinfo  {journal}
					{Phys. Rev. Lett.}\ }\textbf {\bibinfo {volume} {122}},\ \bibinfo {pages}
				{063602} (\bibinfo {year} {2019})}\BibitemShut {NoStop}%
			\bibitem [{\citenamefont {Dufour}\ \emph {et~al.}(2020)\citenamefont {Dufour},
				\citenamefont {Br{\"u}nner}, \citenamefont {Rodr{\'\i}guez},\ and\
				\citenamefont {Buchleitner}}]{dufour2020many}%
			\BibitemOpen
			\bibfield  {author} {\bibinfo {author} {\bibfnamefont {G.}~\bibnamefont
					{Dufour}}, \bibinfo {author} {\bibfnamefont {T.}~\bibnamefont {Br{\"u}nner}},
				\bibinfo {author} {\bibfnamefont {A.}~\bibnamefont {Rodr{\'\i}guez}}, \ and\
				\bibinfo {author} {\bibfnamefont {A.}~\bibnamefont {Buchleitner}},\ }\href
			{\doibase 10.1088/1367-2630/abb7a3} {\bibfield  {journal} {\bibinfo
					{journal} {New Journal of Physics}\ }\textbf {\bibinfo {volume} {22}},\
				\bibinfo {pages} {103006} (\bibinfo {year} {2020})},\ \Eprint
			{http://arxiv.org/abs/2005.07239} {2005.07239} \BibitemShut {NoStop}%
			\bibitem [{\citenamefont {Brunner}\ \emph {et~al.}(2023)\citenamefont
				{Brunner}, \citenamefont {Pausch}, \citenamefont {Carnio}, \citenamefont
				{Dufour}, \citenamefont {Rodr{\'\i}guez},\ and\ \citenamefont
				{Buchleitner}}]{brunner2023many}%
			\BibitemOpen
			\bibfield  {author} {\bibinfo {author} {\bibfnamefont {E.}~\bibnamefont
					{Brunner}}, \bibinfo {author} {\bibfnamefont {L.}~\bibnamefont {Pausch}},
				\bibinfo {author} {\bibfnamefont {E.~G.}\ \bibnamefont {Carnio}}, \bibinfo
				{author} {\bibfnamefont {G.}~\bibnamefont {Dufour}}, \bibinfo {author}
				{\bibfnamefont {A.}~\bibnamefont {Rodr{\'\i}guez}}, \ and\ \bibinfo {author}
				{\bibfnamefont {A.}~\bibnamefont {Buchleitner}},\ }\href {\doibase
				10.1103/PhysRevLett.130.080401} {\bibfield  {journal} {\bibinfo  {journal}
					{Physical Review Letters}\ }\textbf {\bibinfo {volume} {130}},\ \bibinfo
				{pages} {080401} (\bibinfo {year} {2023})},\ \Eprint
			{http://arxiv.org/abs/2209.06871} {2209.06871} \BibitemShut {NoStop}%
			\bibitem [{\citenamefont {Tichy}\ \emph {et~al.}(2012)\citenamefont {Tichy},
				\citenamefont {Tiersch}, \citenamefont {Mintert},\ and\ \citenamefont
				{Buchleitner}}]{tichy2012many}%
			\BibitemOpen
			\bibfield  {author} {\bibinfo {author} {\bibfnamefont {M.~C.}\ \bibnamefont
					{Tichy}}, \bibinfo {author} {\bibfnamefont {M.}~\bibnamefont {Tiersch}},
				\bibinfo {author} {\bibfnamefont {F.}~\bibnamefont {Mintert}}, \ and\
				\bibinfo {author} {\bibfnamefont {A.}~\bibnamefont {Buchleitner}},\ }\href
			{\doibase 10.1088/1367-2630/14/9/093015} {\bibfield  {journal} {\bibinfo
					{journal} {New Journal of Physics}\ }\textbf {\bibinfo {volume} {14}},\
				\bibinfo {pages} {093015} (\bibinfo {year} {2012})},\ \Eprint
			{http://arxiv.org/abs/1204.5588} {1204.5588} \BibitemShut {NoStop}%
			\bibitem [{\citenamefont {Renner}(2005)}]{Renner2005}%
			\BibitemOpen
			\bibfield  {author} {\bibinfo {author} {\bibfnamefont {R.}~\bibnamefont
					{Renner}},\ }\emph {\bibinfo {title} {Security of Quantum Key
					Distribution}},\ \href {https://arxiv.org/abs/quant-ph/0512258} {\bibinfo
				{type} {Phd thesis}},\ \bibinfo  {school} {ETH Zurich} (\bibinfo {year}
			{2005})\BibitemShut {NoStop}%
			\bibitem [{\citenamefont {T\'oth}\ \emph {et~al.}(2010)\citenamefont {T\'oth},
				\citenamefont {Wieczorek}, \citenamefont {Gross}, \citenamefont {Krischek},
				\citenamefont {Schwemmer},\ and\ \citenamefont {Weinfurter}}]{Toth2010}%
			\BibitemOpen
			\bibfield  {author} {\bibinfo {author} {\bibfnamefont {G.}~\bibnamefont
					{T\'oth}}, \bibinfo {author} {\bibfnamefont {W.}~\bibnamefont {Wieczorek}},
				\bibinfo {author} {\bibfnamefont {D.}~\bibnamefont {Gross}}, \bibinfo
				{author} {\bibfnamefont {R.}~\bibnamefont {Krischek}}, \bibinfo {author}
				{\bibfnamefont {C.}~\bibnamefont {Schwemmer}}, \ and\ \bibinfo {author}
				{\bibfnamefont {H.}~\bibnamefont {Weinfurter}},\ }\href {\doibase
				10.1103/PhysRevLett.105.250403} {\bibfield  {journal} {\bibinfo  {journal}
					{Phys. Rev. Lett.}\ }\textbf {\bibinfo {volume} {105}},\ \bibinfo {pages}
				{250403} (\bibinfo {year} {2010})}\BibitemShut {NoStop}%
			\bibitem [{\citenamefont {Tichy}(2015)}]{Tichy2015}%
			\BibitemOpen
			\bibfield  {author} {\bibinfo {author} {\bibfnamefont {M.~C.}\ \bibnamefont
					{Tichy}},\ }\href {\doibase https://doi.org/10.1103/PhysRevA.91.022316}
			{\bibfield  {journal} {\bibinfo  {journal} {Phys. Rev. A}\ }\textbf {\bibinfo
					{volume} {91}},\ \bibinfo {pages} {022316} (\bibinfo {year}
				{2015})}\BibitemShut {NoStop}%
			\bibitem [{\citenamefont
				{Shchesnovich}(2015{\natexlab{b}})}]{Shchesnovich2015trace}%
			\BibitemOpen
			\bibfield  {author} {\bibinfo {author} {\bibfnamefont {V.~S.}\ \bibnamefont
					{Shchesnovich}},\ }\href {\doibase 10.1103/PhysRevA.91.063842} {\bibfield
				{journal} {\bibinfo  {journal} {Phys. Rev. A}\ }\textbf {\bibinfo {volume}
					{91}},\ \bibinfo {pages} {063842} (\bibinfo {year}
				{2015}{\natexlab{b}})}\BibitemShut {NoStop}%
			\bibitem [{\citenamefont {Dittel}\ \emph {et~al.}(2021)\citenamefont {Dittel},
				\citenamefont {Dufour}, \citenamefont {Weihs},\ and\ \citenamefont
				{Buchleitner}}]{Dittel2021}%
			\BibitemOpen
			\bibfield  {author} {\bibinfo {author} {\bibfnamefont {C.}~\bibnamefont
					{Dittel}}, \bibinfo {author} {\bibfnamefont {G.}~\bibnamefont {Dufour}},
				\bibinfo {author} {\bibfnamefont {G.}~\bibnamefont {Weihs}}, \ and\ \bibinfo
				{author} {\bibfnamefont {A.}~\bibnamefont {Buchleitner}},\ }\href {\doibase
				10.1103/PhysRevX.11.031041} {\bibfield  {journal} {\bibinfo  {journal} {Phys.
						Rev. X}\ }\textbf {\bibinfo {volume} {11}},\ \bibinfo {pages} {031041}
				(\bibinfo {year} {2021})}\BibitemShut {NoStop}%
			\bibitem [{\citenamefont {Dittel}(2019)}]{Dittel2019}%
			\BibitemOpen
			\bibfield  {author} {\bibinfo {author} {\bibfnamefont {C.}~\bibnamefont
					{Dittel}},\ }\emph {\bibinfo {title} {About the interference of many
					particles}},\ \href {https://diglib.uibk.ac.at/urn:nbn:at:at-ubi:1-47210}
			{\bibinfo {type} {Phd thesis}},\ \bibinfo  {school} {Universit{\"a}t
				Innsbruck} (\bibinfo {year} {2019})\BibitemShut {NoStop}%
			\bibitem [{\citenamefont {Karczewski}\ \emph {et~al.}(2019)\citenamefont
				{Karczewski}, \citenamefont {Pisarczyk},\ and\ \citenamefont
				{Kurzy\ifmmode~\acute{n}\else \'{n}\fi{}ski}}]{Karczewski2019}%
			\BibitemOpen
			\bibfield  {author} {\bibinfo {author} {\bibfnamefont {M.}~\bibnamefont
					{Karczewski}}, \bibinfo {author} {\bibfnamefont {R.}~\bibnamefont
					{Pisarczyk}}, \ and\ \bibinfo {author} {\bibfnamefont {P.}~\bibnamefont
					{Kurzy\ifmmode~\acute{n}\else \'{n}\fi{}ski}},\ }\href {\doibase
				10.1103/PhysRevA.99.042102} {\bibfield  {journal} {\bibinfo  {journal} {Phys.
						Rev. A}\ }\textbf {\bibinfo {volume} {99}},\ \bibinfo {pages} {042102}
				(\bibinfo {year} {2019})}\BibitemShut {NoStop}%
			\bibitem [{\citenamefont {Walschaers}(2020)}]{Walschaers2020}%
			\BibitemOpen
			\bibfield  {author} {\bibinfo {author} {\bibfnamefont {M.}~\bibnamefont
					{Walschaers}},\ }\href {\doibase 10.1088/1361-6455/ab5c30} {\bibfield
				{journal} {\bibinfo  {journal} {J. Phys. B}\ }\textbf {\bibinfo {volume}
					{53}},\ \bibinfo {pages} {043001} (\bibinfo {year} {2020})}\BibitemShut
			{NoStop}%
			\bibitem [{Note1()}]{Note1}%
			\BibitemOpen
			\bibinfo {note} {Using that $D_{\protect \rm tr}(\sigma , \varrho )=\protect
				\mathrm {tr}[(\sigma - \varrho )_+]$, where $(X)_+$ denotes the positive part
				of $X$ and the fact that, due to Schur-Weyl duality (see App.~\ref
				{app_schurweyl}), for any permutation-invariant state, such as the
				indistinguishable part $\varrho _{\protect \rm ind}\equiv P^{(N)}\varrho
				P^{(N)}/p_N$ of $\rho $, there is no coherence between the symmetric subspace
				and its orthogonal complement, we have $D_{\protect \rm tr}(\varrho
				_{\protect \rm ind},\varrho ) = \protect \mathrm {tr}[\varrho _{\protect \rm
					ind}-P^{(N)} \varrho P^{(N)}]=1-p_N$.}\BibitemShut {Stop}%
			\bibitem [{Note2()}]{Note2}%
			\BibitemOpen
			\bibinfo {note} {Note also that pairwise HOM experiments between all
				particles for states $\varrho $ with exactly 0 or 1 particles in each of the
				modes cannot be used to determine $P_N$.}\BibitemShut {Stop}%
			\bibitem [{\citenamefont {Pont}\ \emph {et~al.}(2022)\citenamefont {Pont},
				\citenamefont {Albiero}, \citenamefont {Thomas}, \citenamefont {Spagnolo},
				\citenamefont {Ceccarelli}, \citenamefont {Corrielli}, \citenamefont
				{Brieussel}, \citenamefont {Somaschi}, \citenamefont {Huet}, \citenamefont
				{Harouri}, \citenamefont {Lema\^{\i}tre}, \citenamefont {Sagnes},
				\citenamefont {Belabas}, \citenamefont {Sciarrino}, \citenamefont {Osellame},
				\citenamefont {Senellart},\ and\ \citenamefont {Crespi}}]{Pont2022}%
			
			
			
			\BibitemOpen
			\bibfield  {author} {\bibinfo {author} {\bibfnamefont {M.}~\bibnamefont
					{Pont}}, \bibinfo {author} {\bibfnamefont {R.}~\bibnamefont {Albiero}},
				\bibinfo {author} {\bibfnamefont {S.~E.}\ \bibnamefont {Thomas}}, \bibinfo
				{author} {\bibfnamefont {N.}~\bibnamefont {Spagnolo}}, \bibinfo {author}
				{\bibfnamefont {F.}~\bibnamefont {Ceccarelli}}, \bibinfo {author}
				{\bibfnamefont {G.}~\bibnamefont {Corrielli}}, \bibinfo {author}
				{\bibfnamefont {A.}~\bibnamefont {Brieussel}}, \bibinfo {author}
				{\bibfnamefont {N.}~\bibnamefont {Somaschi}}, \bibinfo {author}
				{\bibfnamefont {H.}~\bibnamefont {Huet}}, \bibinfo {author} {\bibfnamefont
					{A.}~\bibnamefont {Harouri}}, \bibinfo {author} {\bibfnamefont
					{A.}~\bibnamefont {Lema\^{\i}tre}}, \bibinfo {author} {\bibfnamefont
					{I.}~\bibnamefont {Sagnes}}, \bibinfo {author} {\bibfnamefont
					{N.}~\bibnamefont {Belabas}}, \bibinfo {author} {\bibfnamefont
					{F.}~\bibnamefont {Sciarrino}}, \bibinfo {author} {\bibfnamefont
					{R.}~\bibnamefont {Osellame}}, \bibinfo {author} {\bibfnamefont
					{P.}~\bibnamefont {Senellart}}, \ and\ \bibinfo {author} {\bibfnamefont
					{A.}~\bibnamefont {Crespi}},\ }\href {\doibase 10.1103/PhysRevX.12.031033}
			{\bibfield  {journal} {\bibinfo  {journal} {Phys. Rev. X}\ }\textbf {\bibinfo
					{volume} {12}},\ \bibinfo {pages} {031033} (\bibinfo {year}
				{2022})}\BibitemShut {NoStop}%
			\bibitem [{\citenamefont {Brunner}(2019)}]{brunner2020many}%
			\BibitemOpen
			\bibfield  {author} {\bibinfo {author} {\bibfnamefont {E.}~\bibnamefont
					{Brunner}},\ }\emph {\bibinfo {title} {Many-body interference, partial distinguishability and entanglement}},\ \href {10.6094/UNIFR/166467} {\bibinfo
				{type} {MSc thesis}},\ \bibinfo  {school} {Albert-Ludwigs-Universit\"at Freiburg} (\bibinfo {year}
			{2019})\BibitemShut {NoStop}%
			\bibitem [{Note3()}]{Note3}%
			\BibitemOpen
			\bibinfo {note} {Note that in Refs.~\cite {brunner2020many,Brunner2022},
				$p_2$ has been used to derive an upper bound on $p_N$ and it has been related
				to another notion of indistinguishability~\cite {Brod2019}. Note that these
				results only apply to a particular subset of so called externally separable
				states. For externally separable states (with fixed total particle number)
				the expectation values $p_k$ agree with the mean coherence, which is the
				quantity studied in Refs.~\cite {brunner2020many,Brunner2022}. For more
				details and the definitions of external separability and the mean coherence
				we refer the reader to Refs.~\cite {brunner2020many,Brunner2022} for more
				details)}\BibitemShut {NoStop}%
			\bibitem [{Note4()}]{Note4}%
			\BibitemOpen
			\bibinfo {note} {For particular subsets of input states, $p_2$ can even be
				determined with less measurements. It is straightforward to show that for
				input states with exactly 0 or 1 particles in each external mode, the
				measurement outcomes from all $N(N-1)/2$ pairwise HOM experiments between all
				particles can be used to compute $p_2$.}\BibitemShut {Stop}%
			\bibitem [{Note5()}]{Note5}%
			\BibitemOpen
			\bibinfo {note} {An instance of a measure which does not only consider $p_N$
				is the purity of $\varrho _{ex}$ as studied in Ref.~\cite
				{brunner2023many}.}\BibitemShut {Stop}%
			\bibitem [{\citenamefont {Kirk}\ \emph {et~al.}(2022)\citenamefont {Kirk},
				\citenamefont {Cotler}, \citenamefont {Huang},\ and\ \citenamefont
				{Lukin}}]{Kirk2022}%
			\BibitemOpen
			\bibfield  {author} {\bibinfo {author} {\bibfnamefont {K.~V.}\ \bibnamefont
					{Kirk}}, \bibinfo {author} {\bibfnamefont {J.}~\bibnamefont {Cotler}},
				\bibinfo {author} {\bibfnamefont {H.-Y.}\ \bibnamefont {Huang}}, \ and\
				\bibinfo {author} {\bibfnamefont {M.~D.}\ \bibnamefont {Lukin}},\ }\href@noop
			{} {\  (\bibinfo {year} {2022})},\ \Eprint {http://arxiv.org/abs/2212.06084}
			{arXiv:2212.06084 [quant-ph]} \BibitemShut {NoStop}%
			\bibitem [{\citenamefont {Huang}\ \emph {et~al.}(2020)\citenamefont {Huang},
				\citenamefont {Kueng},\ and\ \citenamefont {Preskill}}]{huang2020predicting}%
			\BibitemOpen
			\bibfield  {author} {\bibinfo {author} {\bibfnamefont {H.-Y.}\ \bibnamefont
					{Huang}}, \bibinfo {author} {\bibfnamefont {R.}~\bibnamefont {Kueng}}, \ and\
				\bibinfo {author} {\bibfnamefont {J.}~\bibnamefont {Preskill}},\ }\href
			{\doibase https://doi.org/10.1038/s41567-020-0932-7} {\bibfield  {journal}
				{\bibinfo  {journal} {Nature Physics}\ }\textbf {\bibinfo {volume} {16}},\
				\bibinfo {pages} {1050} (\bibinfo {year} {2020})}\BibitemShut {NoStop}%
			\bibitem [{\citenamefont {Elben}\ \emph {et~al.}(2020)\citenamefont {Elben},
				\citenamefont {Kueng}, \citenamefont {Huang}, \citenamefont {van Bijnen},
				\citenamefont {Kokail}, \citenamefont {Dalmonte}, \citenamefont {Calabrese},
				\citenamefont {Kraus}, \citenamefont {Preskill}, \citenamefont {Zoller} \emph
				{et~al.}}]{elben2020mixed}%
			\BibitemOpen
			\bibfield  {author} {\bibinfo {author} {\bibfnamefont {A.}~\bibnamefont
					{Elben}}, \bibinfo {author} {\bibfnamefont {R.}~\bibnamefont {Kueng}},
				\bibinfo {author} {\bibfnamefont {H.-Y.~R.}\ \bibnamefont {Huang}}, \bibinfo
				{author} {\bibfnamefont {R.}~\bibnamefont {van Bijnen}}, \bibinfo {author}
				{\bibfnamefont {C.}~\bibnamefont {Kokail}}, \bibinfo {author} {\bibfnamefont
					{M.}~\bibnamefont {Dalmonte}}, \bibinfo {author} {\bibfnamefont
					{P.}~\bibnamefont {Calabrese}}, \bibinfo {author} {\bibfnamefont
					{B.}~\bibnamefont {Kraus}}, \bibinfo {author} {\bibfnamefont
					{J.}~\bibnamefont {Preskill}}, \bibinfo {author} {\bibfnamefont
					{P.}~\bibnamefont {Zoller}},  \emph {et~al.},\ }\href {\doibase
				https://doi.org/10.1103/PhysRevLett.125.200501} {\bibfield  {journal}
				{\bibinfo  {journal} {Physical Review Letters}\ }\textbf {\bibinfo {volume}
					{125}},\ \bibinfo {pages} {200501} (\bibinfo {year} {2020})}\BibitemShut
			{NoStop}%
			\bibitem [{\citenamefont {Simon}(1996)}]{Simon1996}%
			\BibitemOpen
			\bibfield  {author} {\bibinfo {author} {\bibfnamefont {B.}~\bibnamefont
					{Simon}},\ }\href {\doibase https://doi.org/10.1090/gsm/010} {\emph {\bibinfo
					{title} {Representations of Finite and Compact Groups}}},\ \bibinfo {series}
			{Graduate Studies in Mathematics}, Vol.~\bibinfo {volume} {10}\ (\bibinfo
			{publisher} {American Mathematical Society},\ \bibinfo {year}
			{1996})\BibitemShut {NoStop}%
			\bibitem [{\citenamefont {Fulton}\ and\ \citenamefont
				{Harris}(1991)}]{fulton1991representation}%
			\BibitemOpen
			\bibfield  {author} {\bibinfo {author} {\bibfnamefont {W.}~\bibnamefont
					{Fulton}}\ and\ \bibinfo {author} {\bibfnamefont {J.}~\bibnamefont
					{Harris}},\ }\href {\doibase https://doi.org/10.1007/978-1-4612-0979-9}
			{\emph {\bibinfo {title} {Representation Theory: A First Course}}},\ Graduate
			Texts in Mathematics\ (\bibinfo  {publisher} {Springer New York},\ \bibinfo
			{year} {1991})\BibitemShut {NoStop}%
			\bibitem [{\citenamefont {Tung}(1985)}]{tung1985group}%
			\BibitemOpen
			\bibfield  {author} {\bibinfo {author} {\bibfnamefont {W.}~\bibnamefont
					{Tung}},\ }\href {\doibase https://doi.org/10.1142/0097} {\emph {\bibinfo
					{title} {Group Theory in Physics}}}\ (\bibinfo  {publisher} {World
				Scientific},\ \bibinfo {year} {1985})\BibitemShut {NoStop}%
			\bibitem [{\citenamefont {i~Brugu{\'e}s}(2016)}]{i2016characterizing}%
			\BibitemOpen
			\bibfield  {author} {\bibinfo {author} {\bibfnamefont {J.~T.}\ \bibnamefont
					{i~Brugu{\'e}s}},\ }\href {\doibase 10.1007/978-3-319-49571-2_1} {\emph
				{\bibinfo {title} {Characterizing entanglement and quantum correlations
						constrained by symmetry}}}\ (\bibinfo  {publisher} {Springer},\ \bibinfo
			{year} {2016})\BibitemShut {NoStop}%
			\bibitem [{\citenamefont {Schwemmer}(2015)}]{Schwemmer2015}%
			\BibitemOpen
			\bibfield  {author} {\bibinfo {author} {\bibfnamefont {C.}~\bibnamefont
					{Schwemmer}},\ }\emph {\bibinfo {title} {Efficient tomography and
					entanglement detection of multiphoton states}},\ \href {\doibase \doibase
				10.5282/edoc.19113} {\bibinfo {type} {Phd thesis}},\ \bibinfo  {school}
			{Ludwig-Maximilians-Universit{\"a}t M{\"u}nchen} (\bibinfo {year}
			{2015})\BibitemShut {NoStop}%
			\bibitem [{\citenamefont {Moroder}\ \emph {et~al.}(2012)\citenamefont
				{Moroder}, \citenamefont {Hyllus}, \citenamefont {Tóth}, \citenamefont
				{Schwemmer}, \citenamefont {Niggebaum}, \citenamefont {Gaile}, \citenamefont
				{Gühne},\ and\ \citenamefont {Weinfurter}}]{Moroder2012}%
			\BibitemOpen
			\bibfield  {author} {\bibinfo {author} {\bibfnamefont {T.}~\bibnamefont
					{Moroder}}, \bibinfo {author} {\bibfnamefont {P.}~\bibnamefont {Hyllus}},
				\bibinfo {author} {\bibfnamefont {G.}~\bibnamefont {Tóth}}, \bibinfo
				{author} {\bibfnamefont {C.}~\bibnamefont {Schwemmer}}, \bibinfo {author}
				{\bibfnamefont {A.}~\bibnamefont {Niggebaum}}, \bibinfo {author}
				{\bibfnamefont {S.}~\bibnamefont {Gaile}}, \bibinfo {author} {\bibfnamefont
					{O.}~\bibnamefont {Gühne}}, \ and\ \bibinfo {author} {\bibfnamefont
					{H.}~\bibnamefont {Weinfurter}},\ }\href {\doibase
				10.1088/1367-2630/14/10/105001} {\bibfield  {journal} {\bibinfo  {journal}
					{New J. Phys.}\ }\textbf {\bibinfo {volume} {14}},\ \bibinfo {pages} {105001}
				(\bibinfo {year} {2012})}\BibitemShut {NoStop}%
			\bibitem [{\citenamefont {Rothaus}\ and\ \citenamefont
				{Thompson}(1966)}]{Rothaus1966}%
			\BibitemOpen
			\bibfield  {author} {\bibinfo {author} {\bibfnamefont {O.}~\bibnamefont
					{Rothaus}}\ and\ \bibinfo {author} {\bibfnamefont {J.~G.}\ \bibnamefont
					{Thompson}},\ }\href {\doibase pjm/1102994418} {\bibfield  {journal}
				{\bibinfo  {journal} {Pac. J. Math}\ }\textbf {\bibinfo {volume} {18}},\
				\bibinfo {pages} {175} (\bibinfo {year} {1966})}\BibitemShut {NoStop}%
			\bibitem [{\citenamefont {Diaconis}(1988)}]{Diaconis1988}%
			\BibitemOpen
			\bibfield  {author} {\bibinfo {author} {\bibfnamefont {P.}~\bibnamefont
					{Diaconis}},\ }\href {http://www.jstor.org/stable/4355560} {\bibfield
				{journal} {\bibinfo  {journal} {Lect. Notes-Monogr. Ser.}\ }\textbf {\bibinfo
					{volume} {11}} (\bibinfo {year} {1988})}\BibitemShut {NoStop}%
			\bibitem [{Note6()}]{Note6}%
			\BibitemOpen
			\bibinfo {note} {Recall that the characters of inequivalent irreps are
				mutually orthogonal. Hence, the character of a representation defines
				uniquely its decomposition into irreps.}\BibitemShut {Stop}%
			\bibitem [{\citenamefont {Ou}(2006)}]{Ou2006}%
			\BibitemOpen
			\bibfield  {author} {\bibinfo {author} {\bibfnamefont {Z.~Y.}\ \bibnamefont
					{Ou}},\ }\href {\doibase 10.1103/PhysRevA.74.063808} {\bibfield  {journal}
				{\bibinfo  {journal} {Phys. Rev. A}\ }\textbf {\bibinfo {volume} {74}},\
				\bibinfo {pages} {063808} (\bibinfo {year} {2006})}\BibitemShut {NoStop}%
			\bibitem [{\citenamefont {Dicke}(1954)}]{Dicke1954}%
			\BibitemOpen
			\bibfield  {author} {\bibinfo {author} {\bibfnamefont {R.~H.}\ \bibnamefont
					{Dicke}},\ }\href {\doibase 10.1103/PhysRev.93.99} {\bibfield  {journal}
				{\bibinfo  {journal} {Phys. Rev.}\ }\textbf {\bibinfo {volume} {93}},\
				\bibinfo {pages} {99} (\bibinfo {year} {1954})}\BibitemShut {NoStop}%
			\bibitem [{\citenamefont {Benatti}\ \emph {et~al.}(2020)\citenamefont
				{Benatti}, \citenamefont {Floreanini}, \citenamefont {Franchini},\ and\
				\citenamefont {Marzolino}}]{benatti2020entanglement}%
			\BibitemOpen
			\bibfield  {author} {\bibinfo {author} {\bibfnamefont {F.}~\bibnamefont
					{Benatti}}, \bibinfo {author} {\bibfnamefont {R.}~\bibnamefont {Floreanini}},
				\bibinfo {author} {\bibfnamefont {F.}~\bibnamefont {Franchini}}, \ and\
				\bibinfo {author} {\bibfnamefont {U.}~\bibnamefont {Marzolino}},\ }\href
			{\doibase 10.1016/j.physrep.2020.07.003} {\bibfield  {journal} {\bibinfo
					{journal} {Phys. Rep.}\ }\textbf {\bibinfo {volume} {878}},\ \bibinfo {pages}
				{1} (\bibinfo {year} {2020})}\BibitemShut {NoStop}%
			\bibitem [{\citenamefont {Englbrecht}(2023)}]{MatthiasThesis}%
			\BibitemOpen
			\bibfield  {author} {\bibinfo {author} {\bibfnamefont {M.}~\bibnamefont
					{Englbrecht}},\ }\emph {\bibinfo {title} {Entanglement and correlations in multipartite systems}},\  {\bibinfo
				{type} {Phd thesis}},\ \bibinfo  {school} {Universit\"at Innsbruck} (\bibinfo {year}
			{in preparation})\BibitemShut {NoStop}%
		\end{thebibliography}
	\end{document}